\documentclass[amsmath,amssymb,floatfix,
aps,prx,10pt,noshowpacs,twocolumn,superscriptaddress,
]{revtex4-2}

\usepackage{bbm}
\usepackage{comment}
\usepackage{graphicx}
\usepackage{dcolumn}
\usepackage{bm}
\usepackage{hyperref}
\usepackage{physics}
\usepackage[dvipsnames]{xcolor}
\usepackage{standalone}[border=1pt]
\usepackage{tikz}
\usetikzlibrary{decorations.markings}
\usetikzlibrary{arrows.meta}
\usepackage{amsmath,amssymb}
\usepackage{mathtools}
\usepackage[mathscr]{euscript}
\usepackage{float}
\usepackage{booktabs}

\usepackage{enumitem}

\usepackage{scalerel}
\usepackage{stackengine,wasysym}

\newcommand{\oin}{o_{\text{in}}}
\newcommand{\oout}{o_{\text{out}}}
\newcommand{\cc}{\mathscr{C}}
\newcommand{\cre}{c^{\dagger}}
\newcommand{\ani}{c}
\newcommand{\gq}{G^>}
\newcommand{\gl}{G^<}
\newcommand{\gr}{G^R}
\newcommand{\ga}{G^A}

\newcommand{\oo}{\omega}

\newcommand{\norb}{n_{\mathrm{orb}}}

\begin{document}
\title{Real-Frequency Diagrammatics with Pole Representations}
\author{Rayan Farid}
\thanks{These authors contributed equally}
\affiliation{Institute of Theoretical Physics, Faculty of Physics, University of Warsaw, Warsaw, Poland}%
\author{Lei Zhang}
\thanks{These authors contributed equally}
\affiliation{Department of Physics, University of Michigan, Ann Arbor, Michigan 48109, USA}%
\author{Aiman Al-Eryani}%
\affiliation{Institute of Theoretical Physics, Faculty of Physics, University of Warsaw, Warsaw, Poland}%
\author{Agnieszka Ja\.{z}d\.{z}ewska}%
\affiliation{Institute of Theoretical Physics, Faculty of Physics, University of Warsaw, Warsaw, Poland}%
\author{Emanuel Gull}
\affiliation{Institute of Theoretical Physics, Faculty of Physics, University of Warsaw, Warsaw, Poland}%
\affiliation{Department of Physics, University of Michigan, Ann Arbor, Michigan 48109, USA}%

\date{\today}

\begin{abstract}
Finite-temperature many-body theories are typically formulated in imaginary time, where they are amenable to efficient numerical treatment.
However, the extraction of spectral properties from the imaginary axis is difficult and requires numerical analytic continuation, while a direct formulation on the real axis requires prohibitively expensive numerical quadrature.
In this paper, we represent the objects of many-body perturbation theory using systematically improvable pole and moment expansions and develop the corresponding framework for evaluating diagrammatic contributions directly on the real axis. We introduce a controlled recompression procedure that prevents the proliferation of poles under algebraic and diagrammatic operations.
We benchmark the approach on paradigmatic quantum impurity problems and apply it to self-consistent calculations of the uniform electron gas. The resulting spectral quantities converge systematically towards numerical accuracy.
This work provides a practical framework for controlled and systematically improvable calculations of spectral quantities within diagrammatic many-body theories, including self-consistent conserving approaches.
\end{abstract}

\maketitle
\def\thefootnote{*}\footnotetext{These authors contributed equally to this work}\def\thefootnote{\arabic{footnote}}

\section{\label{sec:Into}Introduction}
Quantum statistical mechanics describes the behavior of quantum systems subject to thermal and, potentially, particle-number fluctuations. Finite-temperature field theory provides a powerful framework for calculating their physical properties. Its central objects are Green’s functions, which in the standard finite-temperature many-body formalism are evaluated in imaginary time or at discrete Matsubara frequencies. This representation provides direct access to  statistical observables in equilibrium.

Spectral quantities, such as the single-particle spectral function or the optical conductivity, are closely related to these Matsubara Green's functions and could in principle be extracted via numerical analytical continuation. However, continuation is ill conditioned and in practice may introduce substantial errors that limit the predictive ability of these simulations. How to obtain reliable spectral quantities for interacting quantum systems is therefore a central open question in Green's function theory.

\begin{figure}[t]
    \begin{tikzpicture}[>={Stealth[length=6pt]}, thick]
  \draw[->, thin] (-0.5,0) -- (7,0) node[right] {$\Re\,t$};
  \draw[decoration={markings, mark=at position 0.35 with {\arrow{>}}},
        postaction={decorate}] (0,0.25) -- (6.2,0.25) node[midway, above] {$\mathscr{C}_-$};
  \draw (6.2,0.25) arc (90:-90:0.25);
  \draw[decoration={markings, mark=at position 0.75 with {\arrow{>}}},
        postaction={decorate}] (6.2,-0.25) -- (0,-0.25) node[midway, below] {$\mathscr{C}_+$};
  \fill (1.5,0.25)  circle (1.5pt) node[above] {$t$};
  \fill (4.5,-0.25) circle (1.5pt) node[below] {$t'$};
\end{tikzpicture}
    \caption{Keldysh contour $\cc=\cc_-\cup\cc_+$: the negative
    (chronological) branch $\cc_-$ runs $-\infty\to+\infty$, the positive
    (anti-chronological) branch $\cc_+$ runs back $+\infty\to-\infty$.}
    \label{fig:contour}
\end{figure}
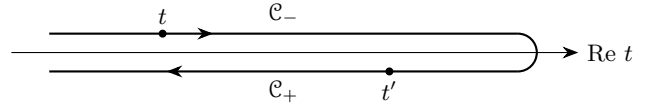

The issue can be overcome in principle by reformulating the equations in terms of equilibrium Keldysh diagrammatics directly on the real axis (Fig.~\ref{fig:contour}). However, the accurate numerical solution of the equations of finite-temperature field theory on a frequency grid is highly non-trivial, and the simultaneous resolution of sharp excitation poles together with broad dispersive features makes the formalism impractical. This problem is particularly acute in self-consistent diagrammatic theories, where the  equations  lead to a proliferation of poles in each iteration.

In this paper, we introduce a different representation of Green's functions in the complex plane, valid simultaneously on both real and Matsubara axes. Our representation is based on an expansion into simple poles in the complex plane. We present a numerical toolkit for performing the operations required in Green's function theory, including frequency summation, Dyson equation solution, diagram contraction, and convolution. Crucially, we present a method for `recompressing' a Green's function with many poles into an equivalent one with fewer poles, thereby avoiding the proliferation of poles plaguing standard real-frequency methods. This recompression step employs a connection to trigonometric moments and the ESPRIT signal processing method. Successful recompression then allows for the routine solution of thermodynamically consistent and conserving self-consistent perturbation theories.

All numerical operations are either accurate to numerical precision or are systematically controlled by an approximation tolerance $\varepsilon$, which can be made arbitrarily small in practice.

\begin{figure*}[tbh]
\includegraphics[width=1.0\linewidth]{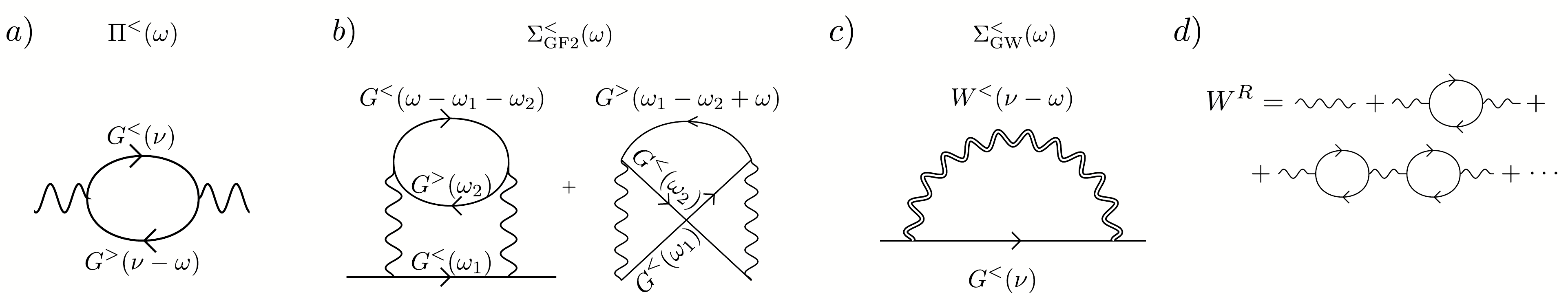}
\caption{Diagrammatic representation of diagrammatic building blocks. \textbf{(a)} Polarization bubble $\Pi^<(t,t')$. \textbf{(b)} Second-order GF2 contributions (direct and exchange terms) to the lesser self-energy $\Sigma_{\mathrm{GF2}}^{<}(t,t')$. \textbf{(c)} Dynamical GW self-energy $\Sigma_{\mathrm{GW}}^{<}(t,t')$ constructed from the Green's function $G^{<}(t,t')$ and the screened interaction $W^{<}(t,t')$. \textbf{(d)} Diagrammatic expansion of the retarded screened interaction $W^R$ in polarization bubbles.
}
\label{fig:diagrams}
\end{figure*}

We demonstrate the power of the method at a few paradigmatic examples, which we solve with the weak coupling perturbation theory methods GF2 (second order Born) and GW, as well as with a dynamical mean field IPT solver. We then show how compact approximations of multi-orbital models with shared poles can be generated by solving a three-orbital model with Slater Kanamori interactions. Finally, as a real-world example for the power of the method we present a solution of the uniform electron gas within self-consistent GW and analyze the satellite features in that theory.

The remainder of this paper is as follows. In Sec.~\ref{sec:Methods} we present the analytical and numerical toolbox for performing real-frequency calculations. This includes the Keldysh formalism, moment and pole representations, the recompression step with ESPRIT, and a summary of diagrammatic operations. In Sec.~\ref{ssec:implementation} we present implementation details for the weak coupling methods used. Sec.~\ref{sec:Results} presents results for quantum impurity problems and the uniform electron gas. Sec.~\ref{sec:discussion} discusses connections to other approaches, further analytic details, and future applications. Finally, Sec.~\ref{sec:conclusions} contains conclusions.

\section{\label{sec:Methods}Methods}
\subsection{\label{ssec:RFD} Diagrammatic Formalism}
To gain direct access to spectral quantities, we formulate diagrammatic perturbative theory in thermal equilibrium
 directly in real-frequency in the Keldysh formalism \cite{Keldysh1964,Stefanucci2013,RammerSmith1986}. The central quantity in this theory
is the contour-ordered Green's function
$G(t_c,t_c')=-i\langle\mathscr{T}_\cc\,\ani(t_c)\cre(t_c')\rangle$, with
time arguments on the closed contour $\cc=\cc_-\cup\cc_+$ of
Fig.~\ref{fig:contour}, running from $t=-\infty$ to $+\infty$ along
$\cc_-$ and back along $\cc_+$. The placement of the two time arguments on
the branches yields four Keldysh components.
Here we primarily work with the lesser, greater, retarded, and
advanced combinations defined as \cite{Stefanucci2013}
\begin{align*}
    &\gl(t,t')=-i\zeta\langle\cre(t')\ani(t)\rangle\,,\quad
    \gq(t,t')=-i\langle\ani(t)\cre(t')\rangle\,,\nonumber\\
    &\gr=\theta(t-t')\big[G^{>}-G^{<}\big]\,,\quad
    \ga=-\theta(t'-t)\big[G^{>}-G^{<}\big]\,.
    \label{eq:keldysh_components}
\end{align*}
Apart from the propagator $G$, other many-body functions such as the polarization bubble $\Pi$, the screened
interaction $W_c$, and the self-energy $\Sigma$ share this structure.  We will refer to any generic many-body function $X$ of this type as a Green's function.

The bosonic ($\zeta=+1$) or fermionic ($\zeta=-1$) statistics of the underlying operators define the Bose and
Fermi-Dirac distribution functions
\begin{align}
 n_{\zeta}(\oo)=(e^{\beta\oo}-\zeta)^{-1}\,,
\end{align}

The time-translation invariance in equilibrium reduces a Green's function $X$ of two times to a single frequency object
$X(\oo)=\int dt \,e^{i\oo t}X(t)$ where $t$ denotes the time difference. We will frequently use the following three properties \cite{Stefanucci2013}: First, components obey
\begin{align}
    X^{>}(\oo)\!-\!X^{<}(\oo)=X^{R}(\oo)\!-\!X^{A}(\oo)
    =:\!\!-2\pi i\,A_X(\oo)\,,
    \label{eq:spectral}
\end{align}
which defines the spectral function
$A_X=-\tfrac{1}{2\pi i}[X^R-(X^R)^\dagger]$. Second, the retarded function
is recovered from the spectral function as the upper boundary value of its
Cauchy-Stieltjes transform $\mathcal{S}$, 
\begin{equation}
X^R(\omega) = \mathcal{S}A_X(\omega+i0^+),\hspace{0.3cm}
\mathcal{S}F(z) := \int \dd{\omega'}\frac{F(\omega')}{z-\omega'}\,,
\label{eq:advanced_retarded_relation}
\end{equation}
which extends to $X^R(z)=\mathcal{S}A_X(z)$ analytic in the entire upper
half plane, with $X^A(z)=[X^R(z^*)]^\dagger$. Third, in equilibrium the
condition $X^{>}(\oo)=\zeta_X e^{\beta\oo}X^{<}(\oo)$ ties the
occupations to the spectrum via the fluctuation-dissipation theorem,
\begin{align}
    X^{<}(\oo)&=-\zeta_X\,n_{\zeta_X}(\oo)\cdot 2\pi i\,A_X(\oo)\,,\nonumber\\
    X^{>}(\oo)&=-\big[1+\zeta_X n_{\zeta_X}(\oo)\big]\,2\pi i\,A_X(\oo)\,.
    \label{eq:FDT}
\end{align}

Any Feynman diagram (Fig.~\ref{fig:diagrams}) decomposes into its Keldysh components, with contour
convolutions resolved by the Langreth rules \cite{Langreth1976,Stefanucci2013} and pointwise contour products
becoming frequency convolutions of $\lessgtr$ components. To demonstrate
the machinery with concrete applications we focus in this paper on two popular conserving approximations, (self-consistent) GW and (self-consistent)
second-order perturbation theory, known as GF2 or second order Born. We emphasize that the formalism presented in this paper remains valid for other diagrammatic schemes, including non-selfconsistent and higher-order perturbation theories.

The GW approximation arises from the self-consistent solution of Hedin's
equations~\cite{Hedin1965,Aryasetiawan1998} upon neglecting vertex corrections and is a popular approximation used in the simulation of materials \cite{Onida2002,Martin2016,Reining2018,Glotze2019}. Its
components take the form 
\begin{align}
    \Pi^{\lessgtr}(\omega)=-i\int\frac{d\nu}{2\pi}
    G^{\lessgtr}(\nu)\,G^{\gtrless}(\nu-\omega)
    \label{eq:pi}
\end{align}
for the particle-hole bubble, shown diagrammatically in Fig.~\ref{fig:diagrams}(a), and
\begin{align}
    W^R(\omega)&=\left[1-v\,\Pi^R(\omega)\right]^{-1}v,\\
    \Pi^R(\omega)&=\mathcal{S}A_\Pi(\omega+i0^+)\,,
    \label{eq:W}
\end{align}
for the screened interaction, whose diagrammatic expansion is shown in
Fig.~\ref{fig:diagrams}(d), where $\Pi^R$ is obtained from
Eqs.~\ref{eq:spectral} and \ref{eq:advanced_retarded_relation} with
$A_\Pi=\tfrac{i}{2\pi}[\Pi^{>}-\Pi^{<}]$ and $v$ denotes the (instantaneous) bare interaction. $W^{\lessgtr}$ follows from Eq.~\ref{eq:FDT} with
$\zeta_X=+1$ applied to the bosonic spectral function of $W$. Since the
$v$ is instantaneous, $v^{\lessgtr}=0$, these coincide with the
components of the correlated screening, $W_c^{\lessgtr}=W^{\lessgtr}$
with $W_c=W-v$. Once $G^{\lessgtr}$ and $W_c^{\lessgtr}$ are obtained,
the correlated part of the self-energy reads
\begin{equation}
    \Sigma^{\lessgtr}_{GW}(\omega)
    =i\int\frac{d\nu}{2\pi}
    G^{\lessgtr}(\nu)\,W_c^{\lessgtr}(\omega-\nu)\,,
    \label{eq:gwse}
\end{equation}
as represented diagrammatically in Fig.~\ref{fig:diagrams}(c). Its retarded component $\Sigma^R_{GW}$ follows from
$\Sigma^{\lessgtr}_{GW}$  with  $\Sigma^R_{GW}=\mathcal{S}A_\Sigma(\oo+i0^+)$ and
$A_\Sigma=\tfrac{i}{2\pi}[\Sigma^{>}_c-\Sigma^{<}_c]$. The bare
interaction contributes the static Hartree-Fock term
\begin{equation}
    \Sigma_{\infty}
    =-i\,v\int\frac{d\nu}{2\pi}\,G^{<}(\nu)\,.
    \label{eq:exchange}
\end{equation}
It is frequency-independent and enters only the retarded component via
$\Sigma^R(\oo)=\Sigma_\infty+\Sigma^R_c(\oo)$, with
$\Sigma_\infty^{\lessgtr}=0$ by $v^{\lessgtr}=0$.
With the notable exception of the uniform electron gas \cite{HolmVonBarth1998}, the self-consistent GW equations are typically not solved self-consistently on the real axis. On the Matsubara axis, powerful implementations of the self-consistent method for solids \cite{Yeh2022,Iskakov2024,Kutepov2010,Pokhilko2024,Yeh2024} form the foundation of real materials embedding schemes \cite{Nilsson2017,Sun2002,KotliarEDMFT2006,BiermannGWDMFT2003,Kananenka2015,Tran2017,Tomczak2012,Zgid2017,RusakovIskakov2019}.

The GF2 or second-order Born method replaces Eq.~\ref{eq:gwse} with a second order perturbative (direct and exchange)  contribution shown diagrammatically in Fig.~\ref{fig:diagrams}(b),
\begin{align}
    \Sigma_{GF2}^{\lessgtr}(\omega)
    =v^2\!\iint\!\frac{d\omega_1d\omega_2}{(2\pi)^2}\,
    G^{\lessgtr}(\omega_1)G^{\gtrless}(\omega_2)
    G^{\lessgtr}(\omega-\omega_1+\omega_2)\,.
    \label{eq:gf2se}
\end{align}
On the Matsubara axis, the method has been successful in understanding the weak correlation physics of molecules \cite{Dahlen2005,Phillips2014,Phillips2015} and of solids with wide gaps \cite{Rusakov2016,Iskakov19}.

The Dyson equation relates $G^R$ to $\Sigma^R$ as
\begin{equation}
    G^R(\omega)=\Big[[G_0^R(\omega)]^{-1}-\Sigma^R(\omega)\Big]^{-1}\,,
    \label{eq:dyson}
\end{equation}
and Eq.~\ref{eq:FDT} relates $G^R$ to $G^{\lessgtr}$.

Eqs.~\ref{eq:pi} through \ref{eq:dyson} form a self-consistent set of equations. In practice, self-consistency is achieved iteratively, starting from an initial guess (such as the Hartree Fock solution). Both approximations are thermodynamically consistent and conserving \cite{LuttingerWard1960,BaymKadanoff1961}. 

\begin{table*}[t]
\centering
\begin{tabular}{@{}llll@{}}
\toprule
Operation & Definition & Output poles / weights & Cost \\
\midrule
Linear comb. & $a\mathbf{J}+b\mathbf{H}$ & $\{\xi^J_j\}\cup\{\xi^H_k\}$;\; $\{aA^J_j\}\cup\{bA^H_k\}$ & $O(M_J{+}M_H)$ \\[2pt]
Adjoint & $\mathbf{J}^{\dagger}$ & $\{(\xi^J_j)^{*}\}$;\; $\{(A^J_j)^{\dagger}\}$ & $O(M_J)$ \\[2pt]
Inversion & $\mathcal{I}\mathbf{J}:\;J(z)\mapsto J(-z)$ & $\{-\xi^J_j\}$;\; $\{-A^J_j\}$ & $O(M_J)$ \\[2pt]
Product & $(\mathbf{J}\mathbf{H})(z)=J(z)H(z)$ & $\{\xi^J_j\}\cup\{\xi^H_k\}$;\; weights by partial fractions, Eq.~\eqref{eq:prodweights} & $O(M_JM_H)$ \\[2pt]
Convolution & $(\mathbf{J}\ast\mathbf{H})(\omega)=\int\!\frac{d\omega'}{2\pi}J(\omega')H(\omega-\omega')$ & $\{\xi^J_j+\xi^H_k\}_{jk}$;\; $\left\{iA^J_jA^H_k\big[\Theta(\operatorname{Im}\xi^J_j)-\Theta\big(\operatorname{Im}(-\xi^H_k)\big)\big]\right\}_{jk}$ & $O(M_JM_H)$ \\[2pt]
Integration & $\int\mathbf{J}=\int_{-\infty}^{\infty}\frac{d\omega}{2\pi}J(\omega)$ & scalar: $i\sum_j\Theta(\operatorname{Im}\xi^J_j)A^J_j$ & $O(M_J)$ \\[2pt]
Boundary value & $\mathcal{S}_{+}\mathbf{A}$: retarded from spectral & keep $\operatorname{Im}\xi_j<0$;\; weights $\times(-2\pi i)$ & $O(M_A)$ \\[2pt]
Dressing & $\mathcal{D}\mathbf{K}:\;K\mapsto(\mathbbm{1}-K)^{-1}-\mathbbm{1}$ & eigenval. of companion matrix Eq.~\ref{eq:pole_representation_dressing_companion}~; Eq.~\ref{eq:pole_representation_dressing_D} & $O\big((nM_K)^3\big)$ \\[2pt]
Compression & $\mathcal{C}_{\varepsilon}\mathbf{X}$ (also $\mathcal{C}\mathbf{X}$, $\mathcal{C}_{M}\mathbf{X}$) & conf. map Eq.~\ref{eq:cayley} + moment Eq.~\ref{eq:moment_linearity} + ESPRIT Sec.~\ref{subsec:esprit} & $O(n^2L^3)$ \\
\bottomrule
\end{tabular}
\caption{\label{tab:ops}
Operations on pole representations and their complexity. Each operation returns a pole representation; $M_X$ denote the input sizes and $n$ the orbital dimension ($
n = \norb$ for fermionic objects such as $\Sigma$ or $G$, and $n = \norb^2$ for bosonic ones such as $\Pi$ or $W$ ). The size of the truncated moment representation is denoted by $L$.}
\end{table*}

\subsection{\label{ssec:Map}Moment and Pole Representation}
\begin{figure}
    \includegraphics[width=1.0\linewidth]{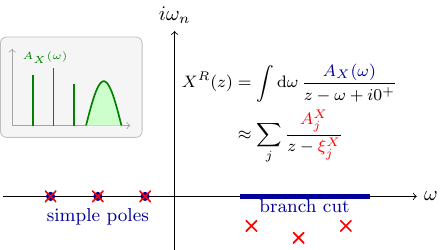}
    \caption{Analytic structure of the spectral function. Blue dots and line: simple poles and branch cuts of $\mathcal{S}A_X(z)$, representing  discrete and continuous parts of the spectrum. Red crosses: locations of complex poles used to approximate $\mathcal{S}A_X(z)$ in the upper half plane. Inset: corresponding real-axis spectral function.}
    \label{fig:minimal_pole_demo}
\end{figure}
Consider a retarded Green's function $X^R$, which is analytic on the upper half plane. We aim to approximate it as a sum over $M_X$ poles 
\begin{align}
X^R(z) \approx \sum_{j=1}^{M_X}
\frac{A^X_j}{z-\xi_j^X},
\quad \Im\xi_j<0,
\label{eq:minimal_pole_ansatz}
\end{align}
where the weights $A^X_j$ are complex and the poles $\xi^X_j$ lie in the lower half plane (LHP). While an exact spectral decomposition contains poles and branch cuts strictly on the real line, corresponding to discrete and continuous spectra, the approximation Eq.~\ref{eq:minimal_pole_ansatz} emulates these branch cuts by shifting poles into the complex plane off the real line, see Fig.~\ref{fig:minimal_pole_demo}.

For cases when $X^R$ is matrix-valued due to its orbital or spinor indices, the weights $A_j^X$ are taken to also be matrix-valued with the diagonal and off diagonal entries sharing their pole structure. In particular, for applications to a system with $\norb$ orbitals, $A^X_j$ will be $\norb \times \norb$ for fermionic objects such as the electron propagator $G^R$ and the $\Sigma^R$, and $\norb^2\times \norb^2$ for bosonic objects such as the polarization bubble $\Pi^R$ and the screened interaction $W_c^R$, unless these objects are  decomposed further \cite{Yeh2022}. 

Refs.~\cite{ZhangGull2024,ZhangYuGull2024,ZhangEtAl2025} showed that
such pole approximations are accurate, systematically improvable, and, for typical condensed-matter spectral functions, often very compact.
Here, we recapitulate the main steps needed to construct this representation. The poles and residues of Eq.~\eqref{eq:minimal_pole_ansatz} are not
fitted directly. Instead, they are recovered from a truncated moment
representation of the Green's function. For numerical stability, we employ a conformal transform $z\mapsto u$ mapping the real axis onto the unit circle and all LHP poles to the interior of the unit disk:
\begin{align}
u(z)=\frac{z+i\omega_p}{z-i\omega_p},\qquad
z(u)=i\omega_p\,\frac{u+1}{u-1},
\label{eq:cayley}
\end{align}
where $\omega_p\in\mathbb{R}^+$ is a free parameter chosen such that
the mapped poles are well separated (we typically choose $\omega_p$ to be half the bandwidth). The trigonometric
moments~\cite{YingAC2022,YingPoleRecovery2022,ZhangEtAl2025}
\begin{align}
h_k=\frac{1}{2\pi i}\oint_{|u|=1} {X^{R}}'(u)\,u^k\,\dd{u},
\qquad k=0,1,\cdots,
\label{eq:moments_quad}
\end{align}
with ${X^{R}}'=X^R\circ z(u)$, provide a stable \cite{Dong2026} moment representation of the
Green's function.

For functions known only on the real axis, Eq.~\ref{eq:moments_quad} is evaluated by standard quadrature on
the unit circle. Where poles and weights are known, either from a previous pole calculation or from a
free-electron dispersion, Eq.~\eqref{eq:moments_quad} follows directly for $\xi'_j=u(\xi_j)$ as
\begin{align}
h_l=\sum_{j}{A^X_j}'\,({\xi^X_j}')^l,
\qquad
{A^X_j}'=\dv{u}{z}\Big|_{\xi_j}A^X_j.
\label{eq:moments_poles}
\end{align}
Since $|{\xi^X_j}'|<1$, the moments decay with $k$, and their computation
is terminated once $|h_k|\le\varepsilon$ or a maximum number of moments is reached.

The construction of the minimum
number of pole and residue pairs $({\xi_j^X}',{A_j^X}')$ that
reproduces the moments to within a specified precision constitutes a
truncated moment problem~\cite{Akhiezer1965}, which we solve in Sec.~\ref{subsec:esprit}. The inverse conformal mapping
then yields $\xi_j^X=z({\xi_j^X}')$ and
$A_j^X=\dv{z}{u}\big|_{{\xi_j^X}'}{A_j^X}'$, which are the poles and weights that go into the pole representation Eq.~\ref{eq:minimal_pole_ansatz}.

Having obtained a pole representation for a retarded Green's function, symmetry implies that the pole representation for its advanced function $X^A$ will have poles $(\xi^X_j)^*$ and weights $[A^X_j]^\dagger$. Furthermore, Eq.~\ref{eq:spectral} requires that a representation for the spectral function $A_X$ on the real axis has poles $\{ \xi^X_j\} \cup \{(\xi^X_j)^*\}$ and weights $\{A_j/2\pi i \}\cup \{-[A^X_j]^\dagger/2\pi i\}$. 

\emph{Notation.} From here on, given a function $J(z)$ (scalar- or matrix-valued) that admits a pole representation, we denote by boldface $ \mathbf{J} = \{\xi^J_j, A^J_j\}_{j=1}^{M_J}$  the data tuple containing its poles and corresponding weights. The number of pole-weight pairs $M_J$ denotes the size of the representation. The evaluation of $\mathbf{J}$ at $z$ is then defined as
\begin{align}
\mathbf{J}(z) = \sum_{j=1}^{M_J} \frac{A_j^J}{z - \xi_j^J},
\label{eq:pole_rep}
\end{align}
where $\mathbf{J}(z) \approx J(z)$ up to the accuracy of the pole representation. 

\subsection{ESPRIT}\label{subsec:esprit}
Given $N$ moments $h_k$, we seek the minimal number $M$ of nodes and
weights satisfying
\begin{equation}
\Big|h_l-\sum_{l=1}^{M}A'_j\,\xi'^{\,l}_j\Big|\le\varepsilon,
\qquad 0\le l\le N-1,
\label{eq:prony_problem}
\end{equation}
where the modulus denotes the maximum absolute deviation among matrix
elements and $\varepsilon$ denotes a predetermined precision. Eq.~\ref{eq:prony_problem} is a Prony-type approximation problem of Eq.~\ref{eq:moments_poles}.  Available solution methods
include the matrix pencil method~\cite{Hua1990,Sarkar1995} and the
estimation of signal parameters via rotational invariance techniques
(ESPRIT)~\cite{RoyKailath1989ESPRIT,PottsTasche2013}. We adopt ESPRIT, which has
demonstrated superior performance~\cite{Takahashi24},
throughout this work. Numerical details are provided in App.~\ref{app:esprit}. Here we summarize the main steps.

In the case of a scalar Green's function, the moments are arranged into a Hankel matrix
$H_{mk}=h_{m+k}$ of dimension $(N-L)\times(L+1)$. The pencil parameter
$L$ is typically chosen in the range $N/3\le L\le N/2$ to minimize the
variance of the recovered nodes~\cite{Sarkar1995}; in this work we fix
it to $L=2N/5$. Since moments generated by $M$ poles yield a Hankel matrix of rank $M$, a singular value decomposition $H=U\Sigma W$ fixes the pole
count as the smallest $M$ determined by the cutoff $\sigma_{M+1}/\sigma_1\le\varepsilon$, with $\sigma$ the singular values of $H$. The nodes $\xi'_j$ are the eigenvalues of
an $M\times M$ matrix built from the leading right-singular
vectors using the shift invariance of the Hankel structure. The
weights follow from a least-squares solution of the Vandermonde system
$h_k=\sum_j A'_j\,\xi'^{\,k}_j$.

In the multi-orbital case, each matrix moment
$h_k\in\mathbb{C}^{n_{\rm orb}\times n_{\rm orb}}$ is flattened into a
row vector. The Hankel matrix is built from these vectors and has
dimension
$(N-L)\times(L+1)\,n_{\rm orb}^2$~\cite{ZhangYuGull2024,YingPoleRecovery2022}.
The decomposition and the shift invariance act on the moment index
alone such that a single set of shared nodes is obtained for all matrix
elements. These are the shared poles of
Eq.~\eqref{eq:minimal_pole_ansatz}. The matrix-valued weights are
recovered element-wise from the same least-squares system.

\subsection{Pole Calculus}\label{sec:MinipoleOperations}
The primary advantage of the pole representation is that the
diagrammatic operations of Sec.~\ref{ssec:RFD} become closed-form
algebraic operations on the poles and weights,  bypassing the need for frequency grids and avoiding grid discretization errors. In the following, we describe these operations; see Tab.~\ref{tab:ops} for a summary.

\emph{Arithmetics.} Linear operations such as taking the adjoint and the
frequency inversion act element-wise on poles and weights, and a
combination $a\mathbf{J}+b\mathbf{H}$ carries the union of the two pole lists with rescaled
weights.  Non-linear operations follow analytically as well. The pointwise
product of two functions $J(z)H(z)$ that each admit pole representations with
non-degenerate pole sets is again a pole representation  $\mathbf{JH}$ containing the union
of the poles. The weights follow from the partial-fraction identity
\begin{align}
\frac{A^J_j}{z-\xi^J_j}\,\frac{A^H_k}{z-\xi^H_k}
=\frac{A^J_j A^H_k}{\xi^J_j-\xi^H_k}\,
\frac{1}{z-\xi^J_j}
+\frac{A^J_j A^H_k}{\xi^H_k-\xi^J_j}\,
\frac{1}{z-\xi^H_k}.
\label{eq:prodweights}
\end{align}

\emph{Integral operations.} Real-frequency integrals are evaluated with the residue theorem. The
integration contour is closed by a semicircular arc in the UHP/LHP yielding $\pm2\pi i$ times the enclosed
residues, provided the integrand decays faster than $1/|\omega|$ so
that the arc does not contribute. For the frequency convolution
$(J*H)(\omega)=\int\frac{\dd{\nu}}{2\pi}J(\nu)H(\omega-\nu)$, the
integrand is a product of two pole sums and decays as $1/\nu^{2}$.
Closing the contour and summing the residues yields, for $\omega \in \mathbb{R}$,
\begin{align}
(J*H)(\omega)=i\sum_{jk}
\big[\Theta(\Im\xi^J_j)-\Theta(-\Im\xi^H_k)\big]\,
\frac{A^J_j A^H_k}{\omega-\xi^J_j-\xi^H_k},
\label{eq:convolution}
\end{align}
a pole sum on the pairwise sums $\xi^J_j+\xi^H_k$, where the
Heaviside functions record whether each pole lies inside the closed
contour and a term contributes only when two poles lie in opposite half-planes. This specifies the convolution as a map from two pole representations to a pole representation $\mathbf{J}\ast \mathbf{H}$. In a similar manner, integrals over the real line evaluate to
the sum of the UHP residues,
\begin{align}
\int_{-\infty}^{\infty}\frac{\dd{\omega}}{2\pi}\,X(\omega)
=i\sum_{\Im\xi_j>0}A_j,
\label{eq:integral}
\end{align}
which is valid provided $X(z)$ decays faster than $|z|^{-1}$. A sufficient equivalent condition is that $\sum_j A^X_j = 0$. We denote this operation by $\int \mathbf{X}$.

Another integral operator is the Cauchy-Stieltjes upper boundary value $\mathcal{S}X(z + i0^+)$ that recovers a retarded function that is analytic in the UHP from a spectral function. Analogously, we define the operator $\mathcal{S}_+$ which maps a pole representation to the result of $\mathcal{S}\mathbf{X}(z+i0^+)$ obtained by closing the integration contour in the LHP. This is implemented as a masking operation which keeps only poles located in the LHP and multiplies the corresponding weights by $-2\pi i$.

\emph{Pole representation for Fermi and Bose distribution functions.} The construction of the lesser and greater functions from their retarded and advanced companions requires additional discussion. By Eq.~\eqref{eq:FDT} they are products of $A_X$, for which we assume a pole representation exists, with the Fermi and Bose distribution functions $n_\zeta$. To maintain a unified formulation of many-body equations strictly in terms of pole operations, we therefore seek pole representations for $n_\zeta$.

The exact Fermi and Bose distribution functions are generated by an infinite number of equidistant simple poles at the imaginary Matsubara frequencies $i\oo_n$~\cite{matsubara_original_1955}, in terms of which the distribution functions could be expanded. However, finite order truncations of the Matsubara series converge too slowly to be useful. Instead, we employ an approximation to the distribution function which has a convenient pole expansion and is rapidly convergent to the exact result on the real axis~\cite{Ozaki2007, Croy2009, Hu2010,Hu2011}, known as the Pad\'e spectral decomposition (PSD). This approximation is
obtained by truncating the continued-fraction representation of
Ozaki~\cite{Ozaki2007} for $\tilde{n}_\zeta \equiv n_\zeta+\zeta/2$,
\begin{align}
\tilde{n}_{\zeta}(z) \approx \sum_{j=1}^{M_\zeta}\frac{A^{\zeta}_j}{z-\xi^{\zeta}_j},
\label{eq:psd}
\end{align} 
with $M_\zeta=2N$ poles for the Fermi and $M_\zeta=2N{+}1$ for the
Bose function, the latter including the pole at zero. The poles of the resulting PSD lie on the imaginary axis and its residues are purely real, making the representation essentially an optimized Matsubara expansion. 

As the exact $\tilde{n}_\zeta$ functions do not decay to zero in all directions, Eq.~\ref{eq:psd} has a finite validity window which grows quadratically with $N$, such that resolving the distribution function valid on a bandwidth $W$ at inverse
temperature $\beta$ requires $N\propto\sqrt{\beta W}$
poles~\cite{Hu2010}.  Within their validity range, the two expansions converge exponentially with $N$ to the corresponding exact distribution functions. Denoting the PSD pole representation as $\mathbf{\tilde{n}_{\zeta}}$, Eq.~\ref{eq:FDT} can be represented directly in terms of operations on pole representations to obtain $\mathbf{X^{\gtrless}}$. The method works well as long as the support of the spectral function lies within the regime of validity of the Pad\'{e} approximant.

We illustrate results and convergence of this methodology for two representative real-frequency convolutions in Fig.~\ref{fig:I12_benchmark}. The first is a fermionic bubble, $I_1(\Omega) =-\int_{-\infty}^{\infty}\frac{\dd{\omega}}{2\pi}
g_0^<(\omega)g_0^>(\omega-\Omega)$,
where we use a synthetic three-peak spectral function at $\beta=10$. The second is a convolution involving a bosonic spectral function, $I_2(\omega) =
\int_{-\infty}^{\infty}\frac{\dd{\Omega}}{2\pi}
d_0^<(\Omega)g_0^<(\omega-\Omega)$,
for a damped bosonic mode, whose form resembles that of the $GW$ self-energy Eq.~\ref{eq:gwse}. Details are given in App.~\ref{app:pade_spectral}. We find that for both $I_1$ and $I_2$, the error decreases systematically with the number of PSD poles $N$, reaching close to numerical at $N=32$.

The remaining operations are the Dyson-type inversions and compressions, which are the subject of Sec.~\ref{ssec:GeomSeries} and Sec.~\ref{sssec:compression}. 

\subsection{\label{ssec:GeomSeries}Resolvent solution of the Dyson equation}

We consider Dyson-type inversions of the form
\begin{align}
    X(z)=\big[X_0^{-1}(z)-B(z)\big]^{-1},
    \label{eq:resolvent_form}
\end{align}
with $(X_0,B)=(G_0^R,\Sigma^R)$ for the propagator in Eq.~\ref{eq:dyson}  and
$(v,\Pi^R)$ for the screened interaction in Eq.~\ref{eq:W}. The dressed poles are
determined by the zeros of the Dyson denominator,
\begin{align}
    \det\!\left[X_0^{-1}(z)-B(z)\right]=0.
    \label{eq:dyson_denominator}
\end{align}
The inversion could be evaluated pointwise on a frequency grid and
the resulting function subsequently converted to a pole
representation through the quadrature moments of
Eq.~\eqref{eq:moments_quad}. To avoid the quadrature cost and discretization error associated
with this procedure, one can instead construct a frequency-independent
matrix in an enlarged space whose eigenvalues yield the dressed poles
directly. Conceptually, this upfolding strategy can be thought of as the inverse of the Feshbach-Löwdin partitioning technique~\cite{lowden1951}, where auxiliary degrees of freedom are projected out to yield a downfolded, effective frequency-dependent Hamiltonian. By reversing this logic, one obtains the Algorithmic Inversion Method (AIM) introduced in Refs.~\cite{ChiarottiMarzariFerretti2022,ChiarottiFerrettiMarzari2024}. We outline this strategy generalized for matrix-valued objects below.

A pole representation $\mathbf{J}$ in Eq.~\ref{eq:pole_rep}, may be expressed as a matrix resolvent
\begin{align}
\mathbf{J}(z)=C_J(z-D_J)^{-1}B_J,
\label{eq}
\end{align}
where
\begin{align}
    D_J
    &= \operatorname{diag}
       (\xi_1^J \otimes\mathbbm{1},\ldots,\xi_{M_J}^J \otimes\mathbbm{1}),
    \nonumber\\
    C_J
    &= (A_1^J\;\cdots\;A_{M_J}^J),
    \nonumber\\
    B_J
    &= (\mathbbm{1}\;\cdots\;\mathbbm{1})^T . 
    \label{eq:realization}
\end{align}
Here $D_J$ contains the poles, $C_J$ collects the corresponding
weights, and $B_J$ stacks $M_J$ copies of the identity in orbital
space. 

In particular, for the two functions $(G_{0}^{R}, \Sigma^{R}_{c}+\Sigma_{\infty})$ entering the fermionic Dyson equation,
 the corresponding pole representation $\mathbf{G_0}$ and $\mathbf{\Sigma^R_c}$ can be written as
\begin{align}
G_0^R(z)
&=C_0(z-D_0)^{-1}B_0,
&
\Sigma_c^R(z)
&=C_\Sigma(z-D_\Sigma)^{-1}B_\Sigma,
\label{eq:matrixresolvents}
\end{align}

Substituting Eq.~\ref{eq:matrixresolvents} into the Dyson equation and applying the
block-resolvent construction, the dressed Green's function can itself be
written as the projected resolvent
\begin{align}
    G^R(z)
    =
    \begin{pmatrix}C_0&0\end{pmatrix}
    (z-H)^{-1}
    \begin{pmatrix}B_0\\0\end{pmatrix},
    \label{eq:projected_resolvent}
\end{align}
of the upfolded matrix
\begin{align}
    H=
    \begin{pmatrix}
        D_0+B_0\Sigma_\infty C_0 & B_0C_\Sigma \\[3pt]
        B_\Sigma C_0             & D_\Sigma
    \end{pmatrix}.
    \label{eq:upfolded_main}
\end{align}
The eigenvalues of $H$ therefore give the dressed pole positions. Since
$H$ has dimension $n_{\rm orb}(M_0+M_\Sigma)$, there will be at most this
many poles.

Diagonalizing $H=\Psi\,\operatorname{diag}(\eta_j)\Psi^{-1}$, with right and left eigenvectors $\psi_j$ and $\phi_j^T$, and
projecting the spectral decomposition of $(z-H)^{-1}$ gives
\begin{align}
    \mathbf{G^R}(z)
    &=\sum_j\frac{A^G_j}{z-\xi^G_j},
&
    A^G_j
    &=\big(C_0\psi_j^{(0)}\big)
      \big(\phi_j^{(0)T}B_0\big),
    \label{eq:dyson_poles}
\end{align}
where the superscript $(0)$ denotes the components in the $G_0$
sector. Thus $\xi^G_j$ are the dressed pole positions and $A^G_j$ the
corresponding matrix residues.

The screened interaction follows from the same construction. Writing
\begin{align}
    \mathbf{\Pi^R}(z)=C_\Pi(z-D_\Pi)^{-1}B_\Pi
\end{align}
and using that the bare interaction $v$ is frequency independent, the upfolded matrix
reduces to
\begin{align}
    H_W=D_\Pi+B_\Pi v C_\Pi .
    \label{eq:upfolded_W}
\end{align}
Its eigenvalues give the poles of the dynamical part
$W_c^R=W^R-v$, with residues
\begin{align}
    A_j^W=
    \big(vC_\Pi\psi_j\big)
    \big(\phi_j^TB_\Pi v\big).\label{eq:residue_W}
\end{align}
The dynamical part vanishes as $|z|\to\infty$, while the constant
asymptote of $W^R$ remains $v$. 

A unified formulation can be obtained by rewriting Eq.~\ref{eq:dyson} as
\begin{align}
X(z) = \mathcal{D}[K]X_0 + X_0
\end{align}
where we have introduced the \emph{dressing operator} $\mathcal{D}[K] = (\mathbbm{1} - K)^{-1} - \mathbbm{1}$, for $K(z) = X_0(z)B(z)$. Unlike the right hand side of Eq.~\ref{eq:dyson}, the result of this operator decays to zero as $|z| \rightarrow \infty$ whether $X_0$ depends on frequency or not. As such, it maps a pole representation to another pole representation. Its poles and weights are recovered in a similar manner; further details are given in App.~\ref{app:dressing_operator}. 

\subsection{\label{sssec:compression}Compression}

The preceding pole calculus operations grow the pole count at every
step. Each convolution, Eq.~\eqref{eq:convolution}, compounds the
sizes of its inputs to $M_JM_H$ output poles, and each Dyson
resolvent, Eq.~\eqref{eq:dyson_poles}, returns the full enlarged
dimension of $n_{\rm orb}(M_0+M_\Sigma)$ and  $n^2_{\rm orb}(M_\Pi)$ poles for $G^R$ and $W^R$, respectively. Repeated application of diagrammatic operations
within a self-consistent cycle would therefore lead to unbounded pole
proliferation and anything beyond a few iterations would be computationally intractable. This proliferation is avoided by the re-compression of the pole representation after 
every significant convolution and Dyson-type inversion. Since
the poles and weights are available, the trigonometric moments follow
analytically from the residue theorem, Eq.~\eqref{eq:moments_poles},
and ESPRIT retrieves a compact representation of the poles of the retarded function on the LHP. For a generic
representation $\mathbf{X}$, the LHP subset of poles will be mapped to the disk, and the moments are
passed to ESPRIT,
\begin{align}
\mathcal{C}\,\mathbf{X}
=u^{*}\circ\operatorname{ESPRIT}
\Big(\sum_j A^{X'}_j\,(\xi^{X'}_j)^k\Big)_{k},
\label{eq:compression}
\end{align}
where $u^{*}$ denotes the pull-back by the conformal map Eq.~\ref{eq:cayley} to the
$z$-plane. In the optimal re-compression $\mathcal{C}\mathbf{X}$, the retained
rank is set by the numerical noise floor, the plateau at which the
singular values of the moment Hankel matrix stop decaying due to the
error accumulated in the diagrammatic operations. In practice, the
pole number is often restricted instead by a tolerance chosen from
the accuracy of the input data, or by a pole budget set by the
available computational resources. We denote these re-compressions to
a tolerance $\varepsilon$ and to at most $M$ poles by
$\mathcal{C}_\varepsilon \mathbf{X}$ and $\mathcal{C}_M \mathbf{X}$, respectively.  Importantly, because recompression isolates the LHP subset of poles to yield a retarded representation, it is always applied either to pole representations of retarded functions or directly following the action of $\mathcal{S}^+$.

To prevent pole proliferation beyond ESPRIT recompression, we can exploit diagrammatic operations that are linear with respect to the trigonometric moments. Operations such as momentum and diagram summations 
concatenate pole lists and quickly become computationally intractable
under repeated application. Instead, a sum
$X=\sum_\alpha c_\alpha X_\alpha$ can be performed on the trigonometric
moments, which are linear in the function,
\begin{align}
h_k=\sum_\alpha c_\alpha\,h^{(\alpha)}_k
=\sum_{\alpha j} c_\alpha\,
 A'^{(\alpha)}_j\,\xi'^{(\alpha)\,k}_j,
\label{eq:moment_linearity}
\end{align}
followed by a single ESPRIT recovery of the summed representation. 

We emphasize that compression is a crucial step in the formalism presented here and the main step making self-consistent diagrammatics viable.
Without it, the self-consistent GW and GF2 cycles and controlled discrete or stochastic momentum summations could not be carried out grid-free due to the rapid growth of the number of poles.

\begin{figure}[tbh]
    \centering
    \includegraphics[width=0.8\columnwidth]{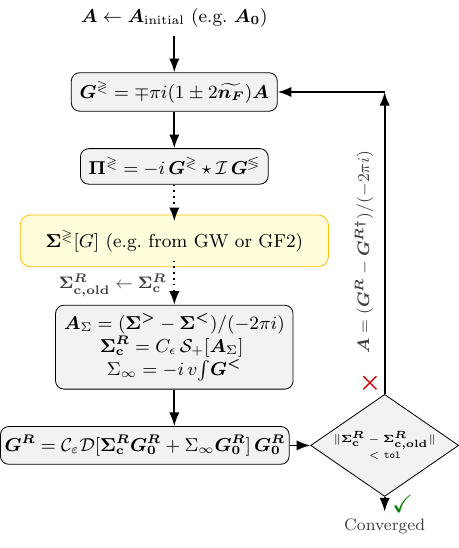}
        \caption{Self-consistent cycle within the pole representation. All operations are algebraic manipulations on poles and weights. The $C_\varepsilon$ operator indicates compression within a desired tolerance $\varepsilon$ which prevents proliferation of poles.}
        \label{fig:selfconsistent}
        
\end{figure}

\section{\label{ssec:implementation}Implementation}
We now show how the pole and moment representations provided by the minimal pole method can be used to formulate self-consistent diagrammatic methods, circumventing the problem of the proliferation of poles expected within these methods. To that end, we present the algebraic equations in the pole representation for two self-consistent frameworks---$GW$, and second-order Green's function perturbation theory (GF2)---all operating directly in real frequency without relying on discretized frequency grids. We emphasize that these methods are used here as examples and that the formalism remains applicable to other diagrammatic methods.

Without loss of generality, we consider a system characterized by a bare retarded Green's function matrix---in the pole representation $\mathbf{G_{0,}^R}_{\oout \oin}$---and a frequency-independent static interaction tensor $U_{(\oout \oin) (\oout' \oin')}$. For clarity, orbital indices are suppressed whenever they can be inferred from context. Grouped indices inside parentheses, such as $(\oout \oin)$, denote composite super-indices; operations written as products of these multi-index quantities implicitly correspond to matrix multiplication over the respective super-indices. Einstein summation convention over repeated orbital indices is implied throughout. In the first iteration, we set $\mathbf{G^R} \leftarrow \mathbf{G_0^R}$.

To begin, the single-particle spectral function matrix is calculated via
\begin{align}
\mathbf{A} = -\frac{1}{2\pi i}\left(\mathbf{G^R} - [\mathbf{G^R}]^\dagger\right), \label{eq:GW_eq1_spectral_function}
\end{align}
linear in time in the number of poles $\mathcal{O}(\norb^2 M_G)$. Next, the greater and lesser Green's functions $\mathbf{G^{\gtrless}}$ are evaluated:
\begin{align}
\mathbf{G^<} &= \pi i(\mathbf{A} + 2\mathbf{\tilde{n}_F}\mathbf{A}), \\
\mathbf{G^>} &= -\pi i(\mathbf{A} - 2\mathbf{\tilde{n}_F}\mathbf{A}),
\end{align}
each is computed in $\mathcal{O}(\norb^2(M_G+M_F)^2)$ time, where $M_F$ is the number of poles in the Pad\'e spectral decomposition (PSD) $\mathbf{\tilde{n}_F}$. 

\begin{figure}[tb]
\includegraphics[width=0.49\columnwidth]{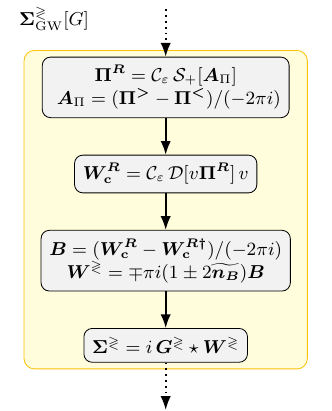}
\includegraphics[width=0.49\columnwidth]{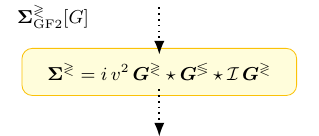}
\caption{Calculation of the greater and lesser self-energies in GW and GF2 within the pole representation.}
\end{figure}

At this stage, the lesser and greater self-energies are calculated, depending on the chosen diagrammatic approximation.

\emph{The $GW$-approximation.}
In $GW$, screening is treated dynamically via the polarization bubble $\mathbf{\Pi^{</>}}$:
\begin{align}
\mathbf{\Pi}^{>}_{(\oout \oin)(\oout' \oin')} &= -i \mathbf{G^{>}}_{\oout \oin'}\ast \mathcal{I}\mathbf{G^{<}}_{\oout' \oin}, \\
\mathbf{\Pi}^{<}_{(\oout \oin)(\oout' \oin')} &= -i\mathbf{G^{<}}_{\oout \oin'}\ast \mathcal{I}\mathbf{G^{>}}_{\oout' \oin},
\end{align}
evaluated in $\mathcal{O}(\norb^4(M_G+M_{F})^2)$ time (note that the orbital scaling changes if decomposed vertices are employed \cite{Yeh2022}). The retarded polarization is obtained via the Cauchy boundary value:
\begin{align}
&\mathbf{\Pi^R} = \mathcal{C}_\varepsilon \mathcal{S}_+[\mathbf{A_\Pi}] \qq{with} \nonumber \\
&\quad\qquad \mathbf{A_\Pi} = -\frac{1}{2\pi i}(\mathbf{\Pi^>} - \mathbf{\Pi^<}),
\end{align}
followed by an ESPRIT compression to a specified error $\varepsilon$ in $\mathcal{O}(\norb^4 L_\Pi^3)$ time, where $L_\Pi$ is the number of moments kept. The correlated retarded screened interaction $\mathbf{W_c^R}$ is then dressed by the polarization
\begin{align}
\mathbf{W^R_c} = \mathcal{C}_\varepsilon \mathcal{D}[v\,\mathbf{\Pi}^R]\,v,
\end{align}
which involves diagonalizing an $\norb^2 M_\Pi \times \norb^2 M_\Pi$ auxiliary matrix in $\mathcal{O}(\norb^6 M_\Pi^3)$ time, where $M_\Pi$ is the number of poles of $\mathbf{\Pi^R}$. From the corresponding bosonic spectral function $\mathbf{B} = \frac{i}{2\pi}(\mathbf{W_c^R} - [\mathbf{W_c^R}]^\dagger)$, the greater and lesser screened interactions are obtained:
\begin{align}
\mathbf{W^<_c} &= -\pi i (-\mathbf{B}+2\mathbf{\tilde{n}_B}\mathbf{B}), \\
\mathbf{W^>_c} &= -\pi i (\mathbf{B} + 2\mathbf{\tilde{n}_B}\mathbf{B}),
\end{align}
and the correlated dynamic self-energy is calculated via:
\begin{align}
\mathbf{\Sigma_{c,}^{>}}_{\oout\oin} = i\mathbf{G}^{>}_{\oin'\oout'}\ast \mathbf{W}^{>}_{c, \, (\oout\oin'),(\oout'\oin)},\\
\mathbf{\Sigma_{c,}^{<}}_{\oout\oin} = i\mathbf{G}^{<}_{\oin'\oout'}\ast \mathbf{W}^{<}_{c, \, (\oout\oin'),(\oout'\oin)}.
\end{align}

\begin{figure*}[tb]
    \centering
    \includegraphics[width=1\linewidth]{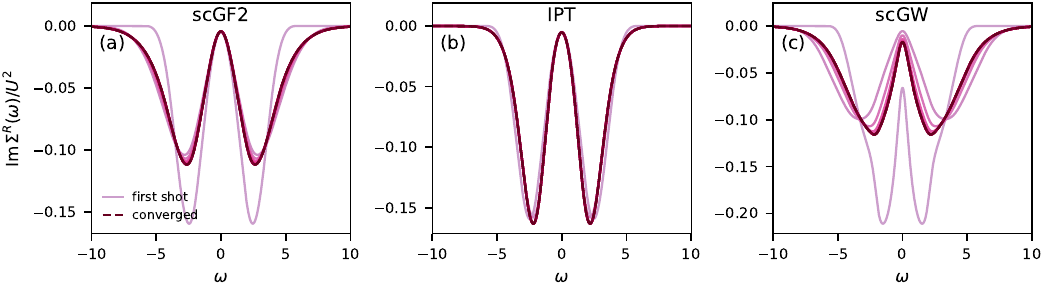}
    
    \caption{Convergence of the self-consistent iterations for the retarded self-energy of an Anderson impurity model describing a local impurity with  $U/t=2$ coupled to a semicircular density of states at inverse temperature $\beta t=10$. First iteration shown in light pink, subsequent iterations indicated by progressively darker shades.  Number of poles  determined by an adaptive recompression scheme  to reach accuracy of $10^{-12}$. Panel (a): GF2 (b): DMFT IPT. (c): GW. 
}
\label{fig:bethe_ImSigR_SelfConsistent}
\end{figure*}

\emph{The second order self-energy.}
In second-order methods, the self-energy is constructed directly from a two-particle second-order bubble without dressing an intermediate interaction $W$. Using the pole convolution operator, the greater and lesser second-order self-energies are formed directly by convolving three Green's function lines \cite{ZgidWithGF2}:
\begin{align}
\mathclap{\mathbf{\Sigma^\gtrless_{c, }}_{\oout\oin}} = i(2&v_{(\oout q)(j k)} v_{(l m)(n \oin)} - v_{(\oout q)(j k)} v_{(n m)(l \oin)}) \nonumber \\ & \times \big( \mathbf{G}^\gtrless_{q n} \ast \mathbf{G}^\lessgtr_{m j} \ast \mathcal{I}\mathbf{G}^\gtrless_{k l} \big),
\end{align}

After determining $\Sigma^>$ and $\Sigma^<$, one calculates the retarded self-energy via
\begin{align}
&\mathbf{\Sigma_{c,}^R} = \mathcal{C}_\varepsilon \mathcal{S}_+\big[\mathbf{A_\Sigma}\big] \qq{with} \nonumber \\
&\qquad \mathbf{A_\Sigma} = -\frac{1}{2\pi i}(\mathbf{\Sigma^{>}_{c}} - \mathbf{\Sigma^{<}_{c}}).
\end{align}

The static Hartree-Fock/exchange contribution is always calculated by
\begin{align}
\Sigma_{\infty,\oout\oin} = -i\,v_{(\oout\oin')(\oout'\oin)}\int \mathbf{G}^<_{\oout'\oin'}.
\end{align}

Finally, the updated retarded Green's function is calculated by applying the dressing operator directly in the pole representation followed by a compression step:
\begin{align}
\mathbf{G^R} = \mathcal{C}_\varepsilon \mathcal{D}[\mathbf{\Sigma_c^R}\mathbf{G_0^R} + \Sigma_\infty \mathbf{G_0^R} ]\ \mathbf{G_0^R}.
\end{align}

Up to this point, this constitutes a single iteration for $GW$ or GF2. Stopping after the first pass corresponds to the single-shot non-iterative approximations ($G_0W_0$ or bare second order perturbation theory (MP2)). Continuing from Eq.~\ref{eq:GW_eq1_spectral_function}, one iterates until the maximal absolute difference in the self-energy $\mathbf{\Sigma^R}$ between consecutive iterations falls below a given tolerance, yielding the fully self-consistent solution ($GW$, $\text{GF2}$).

The  efficient and stable compression scheme (denoted in the equations by $C_\varepsilon$) is crucial to achieve self-consistency: by employing conformal coordinates that map relevant poles into a disk, computing moments with the residue theorem, and applying ESPRIT to retrieve a compact pole representation, the number of poles is kept bounded regardless of the number of diagrammatic convolutions performed.

\section{\label{sec:Results}Results}
We present results for three classes of problems for which our method can be applied straightforwardly. The first is a paradigmatic Anderson impurity model motivated by single-site DMFT of the Hubbard model on the Bethe lattice, where real frequency GF2, $GW$ and IPT impurity solutions are used in the DMFT cycle. The second is an application to a multi-orbital system illustrating shared pole approximations. Finally, we present an application to the uniform electron gas (jellium) within $GW$, which demonstrates the applicability of this method to more complex systems which involve momentum dependence.

\subsection{\label{ssec:SSDMFT} Anderson Impurity Model: GW, GF2 and DMFT IPT}
\emph{Model and setup.} In what follows we illustrate the power of our approach on the Anderson impurity model which describes a local
impurity with on-site interaction $U$ coupled to a bath \cite{Hewson1993}.  We initially employ a semi-circular
density of states, 
\begin{align}
    A(\omega)=\frac{1}{2\pi t^2}\sqrt{4t^2-\omega^2}\,,\qquad |\omega|\le 2t\,.
\end{align}
Throughout we work in the units of $t$ and set $t=1$ and consider the half-filled case at fixed inverse temperature $\beta t=10$. This problem setup is inspired by dynamical mean field calculations, where similar problems emerge in the infinite coordination limit of the Hubbard model on a Bethe lattice \cite{MetznerVollhardt1989,GeorgesKotliar1992,GeorgesKotliar1996}.
The physics of the Anderson impurity model in this context is well known and numerically exact `impurity solver' algorithms exist using diagrammatic \cite{Rubtsov2005,Werner2006,Gull2008,Gull2011,Ge2024}, and other \cite{BullaCostiPruschke2008,Wolf15,Yu2026} techniques, particularly if only Matsubara quantities are desired.
Here, the problem setup and the solution methods are chosen to illustrate the behavior of the pole expansion method as clearly and cleanly as possible. 
\begin{figure*}
    \centering
    \includegraphics[width=0.98\linewidth]{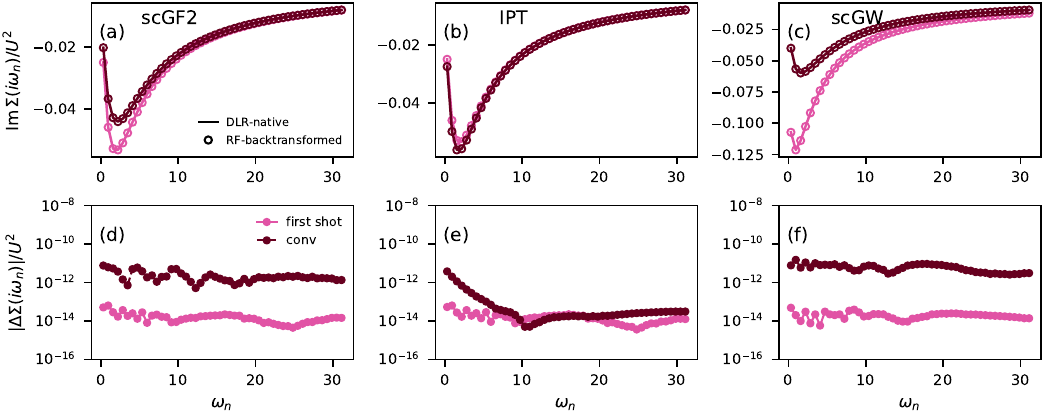}
    \caption{Validation of the real-frequency results against independent Matsubara calculations, for the same choice of parameters as in Fig.~\ref{fig:bethe_ImSigR_SelfConsistent}.
Top row: $\mathrm{Im}\,\Sigma(i\omega_n)$ for (a) GF2, (b) IPT, and (c) GW at the first iteration (light pink) and for the converged
solution (dark purple). Open circles represent the real-axis result back-transformed to
imaginary frequency and solid lines are the Matsubara result with parameters  $\varepsilon = 10^{-14}$ and  $\Lambda = 80$.
Bottom row (d)--(f): absolute deviation
$|\Delta\Sigma(i\omega_n)|/U^2$ between the two calculations. }
\label{fig:bethe_dlr_mastubara}
\end{figure*}
The starting point is a pole representation of the bare spectral function $A(\omega)$. We obtain it by decomposing the semicircular spectral function for which the trigonometric moments
of Eq.~\eqref{eq:moments_quad} are evaluated from the continuous $A(\omega)$ by numerical quadrature on the unit circle with conformal mapping parameter set to $\oo_p=2$, followed by ESPRIT to recover poles and weights. This is the only step of the calculation that involves numerical quadrature; every subsequent step is carried out directly in the pole representation. In the construction of lesser quantities $\mathbf{X^{\lessgtr}}$, the Fermi functions are represented by $N_F=16$ PSD pole pairs. This choice converges the distribution functions to within numerical tolerance over the frequency window in which $A$ has support. The bosonic objects $\Pi$ and $W$ are convolutions of two fermionic propagators and therefore extend over twice the single-particle bandwidth. Correspondingly we use $N_B=36$ to converge bosonic quantities to the same accuracy.

\emph{Self-consistency schemes.}
We consider three schemes. The first two, GW and GF2, are fully self-consistent within their respective approximations where $G^R$ is dressed by the diagrammatic self-energy and fed back into the diagrams as outlined in Sec.~\ref{ssec:implementation}. For these methods, no DMFT self-consistency is employed. The third,  IPT, is the second-order self-energy evaluated on the Weiss propagator of the impurity problem inside a DMFT loop. In IPT, the impurity self-energy is built from the bare propagator $\mathcal{G}_0^R=[\omega+\mu-\Delta^R]^{-1}$
rather than from the dressed $G^R$. The hybridization is updated as ${\Delta^R}=t^2{G^R}$ after every impurity solve. The DMFT self-consistency therefore adds one further Dyson resolvent per iteration on the pole set of $t^2{G^R}$, with $1/(\omega+\mu)$ as the bare term.

Two simplifications follow from the purely local Hubbard interaction. In GF2, the second-order exchange diagram vanishes, since it requires an interaction between equal spins, and the self-energy reduces to the direct term of Eq.~\eqref{eq:gf2se}. In GW, spin conservation at the interaction vertices enforces $\Sigma_{\sigma}=iG_{\sigma}W_{\sigma\sigma}$, so that the equal-spin screening channel $W_{\uparrow \uparrow} =U^2 \Pi/(1-U^2\Pi^2)$ enters the self-energy.

\emph{Convergence of the self-consistent cycle.} Figure~\ref{fig:bethe_ImSigR_SelfConsistent}(a-c) shows the imaginary part of the retarded self-energy for $U/t=2$ at each iteration of the three schemes. The first shot is drawn in light pink and subsequent iterations in progressively darker shades. Convergence is monitored on $\Sigma^R(\omega)$. The iteration is terminated  when the maximal change between consecutive iterations falls below $10^{-12}$. Throughout, the pole number is dynamically adjusted: The compression $\mathcal{C}\,\mathbf{X}$ (Eq.~\ref{eq:compression}) is applied after every convolution and Dyson step with the rank determined by the numerical noise floor of the Hankel moment matrix. IPT converges smoothly and changes little after the first iteration. In GW and GF2, the first iteration overestimates the low-frequency scattering rate, and the feedback of the dressed propagator into the bubble reduces it over the first few iterations before settling.

\emph{Validation against Matsubara calculations.}
Beyond real-frequency spectral functions, our approach naturally provides accurate results along the Matsubara axis. 
To demonstrate this capability against an established benchmark
, we repeat the same three self-consistent cycles on the imaginary axis using the discrete Lehmann representation (DLR)~\cite{DLR_original_PhysRevB.105.235115,Kaye2024_DLR_cpplib}. In imaginary time the diagrams are simple products, $\Pi(\tau)=G(\tau)G(\beta-\tau)$ for the bubble and $\Sigma_{GW}(\tau)=G(\tau)W_c(\tau)$  for the self-energy, while the Dyson equations for $G$ and $W$ are solved on the Matsubara axis. The DLR parameters are $\Lambda= 80$ and $\varepsilon=10^{-14}$. We compare the two calculations on the Matsubara axis rather than on the real axis, since the real-frequency data cannot be obtained from the DLR result without numerical analytic continuation, which is exponentially sensitive to the input and would not allow a comparison on equal footing. On the other hand, back-transformation in complex pole representation is numerically stable. Since $\Sigma^R(z)=\sum_j A^\Sigma_j/(z-\xi^\Sigma_j)$ with $\Im\xi^\Sigma_j<0$ is analytic in the upper half plane, its values at the Matsubara points follow from the substitution $z\to i\omega_n$ for $n>0$.

Fig.~\ref{fig:bethe_dlr_mastubara}(a-c) shows $\Im\Sigma(i\omega_n)$ for the first iteration and the converged solution of all three schemes; (d-f) the absolute deviation between the two calculations. The first iteration agrees to an absolute error of $10^{-13}$, close to the precision of the input. The deviation in self-consistency grows to $10^{-11}$--$10^{-12}$. This loss of precision reflects the accumulated round-off of repeated compressions and resolvent steps and not a systematic error of either method.

\begin{figure}[tb]
    \centering
\includegraphics[width=1\linewidth]{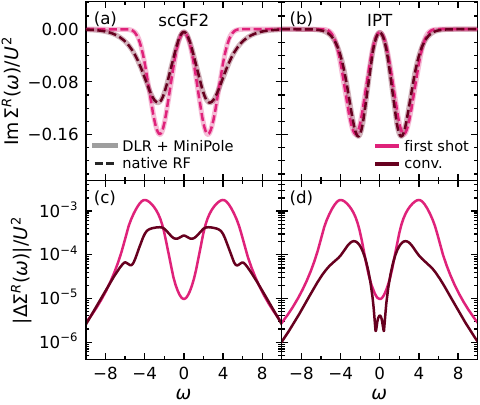}   
    \caption{Comparison of the native real-frequency $\Sigma^{R}(\omega)$ against analytically continued imaginary-axis DLR data (MiniPole AC, $\varepsilon=10^{-11}$). Top row: $\mathrm{Im}\Sigma^{R}(\omega)/U^{2}$ for (a) GF2 and
(b) IPT at the first iteration (light pink) and at the converged
solution (dark purple). Gray lines: continued DLR result.  Dashed lines: native real-frequency
result. Bottom row (c),(d):  absolute deviation
$|\Delta\Sigma^{R}(\omega)|/U^{2}$ as a function of frequency.}
\label{fig:bethe_dlr_3}

\end{figure}

To illustrate the difficulty of comparing spectral functions, we show in Fig.~\ref{fig:bethe_dlr_3} the same comparison on the real axis, where the DLR self-energy is continued to the real axis with the MiniPole analytic continuation at tolerance $\varepsilon=10^{-11}$. 
The analytically continued results struggle to reproduce our real-frequency data beyond $10^{-4}$--$10^{-3}$ accuracy over the whole frequency range. 
\begin{figure}[tbh]
    \centering \includegraphics[width=1\linewidth]{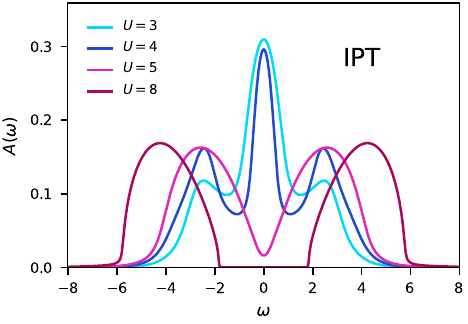}
    \caption{Evolution of the spectral function $A$ on half-filled Bethe lattice for interaction strength $U/t=3,4,5,8$ at a fixed $\beta t=10$. DMFT with IPT solver.}
    \label{fig:bethe_Aevol}
\end{figure}

\emph{Interaction dependence.}
We show the evolution of the spectral function with interaction strength in Fig.~\ref{fig:bethe_Aevol}, for $U/t=3,4,5,8$ at $\beta t=10$, using IPT DMFT~\cite{GeorgesKotliar1992} with the compression tolerance $\varepsilon=10^{-10}$. The plot illustrates the well-known Mott transition in single site DMFT: with increasing $U$, spectral weight is transferred from the Fermi level to the Hubbard bands, and at $U/t=8$ the system is fully gapped.  We note that, near the Mott transition at $U/t=8$, both IPT self-energy and Weiss propagator develops a pole infinitesimally close to real axis at $\omega=0$. While ESPRIT resolves poles close to the real axis to within the specified tolerance $\varepsilon$, the imaginary part of such an isolated pole can fall below this tolerance and the numerical round-off may place it in the upper half plane. Under the mapping of Eq.~\ref{eq:cayley}, such a pole is then discarded in the compression step, leading to numerical instability.  We resolve this by invoking the standard definition of the retarded function:  whenever a pole lies within the compression tolerance $\varepsilon$ of the real axis, its imaginary part is clamped to $-\eta$, with $\eta$ chosen far below all physical scales but above $\varepsilon$ ($\eta=10^{-6}$ here).  We have verified that the results are insensitive to this choice by repeating the calculation for $\eta=10^{-3}$, $10^{-4}$, and $10^{-5}$ for the $U/t =8$ calculation. Alternatively, the regulator can be avoided altogether by choosing a different mapping~\cite{ZhangEtAl2025}, or by treating the atomic propagator exactly and compressing only the remaining continuum.

\begin{figure}[tb]
    \centering
\includegraphics[width=1\columnwidth]{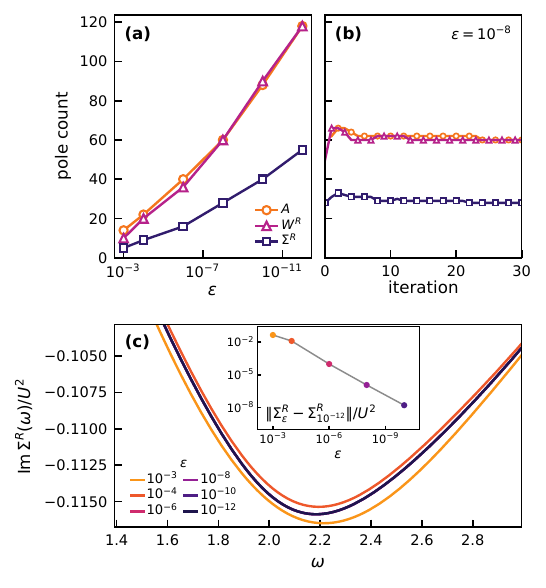} 
    \caption{Pole count of $\Sigma^{R}$, $W^{R}$, and $A$ after the re-compression step within the GW approximation, for the half-filled Bethe lattice at $\beta t = 10$ and $U/t = 2$, (a) as a function of the specified compression tolerance $\varepsilon$ at convergence and (b) as a function of iteration for $\varepsilon = 10^{-8}$. (c) Enlarged view of $\mathrm{Im}\,\Sigma^{R}(\omega)/U^{2}$ from Fig.~\ref{fig:bethe_ImSigR_SelfConsistent}(c) at convergence for tolerances $\varepsilon = 10^{-3}$ to $10^{-12}$. Inset: deviation from the $\varepsilon = 10^{-12}$ result as a function of $\varepsilon$.}
\label{fig:pole_budget}
\end{figure}

\emph{Pole budget.}
So far we chose the number of poles to be set by the numerical noise floor. In practice, given other sources of errors (such as the quality of the diagrammatic approximation itself), lower precision may be sufficient and desirable, since the pole count governs the computational cost. Figure~\ref{fig:pole_budget}(a) shows the number of poles retained for $A$, $W^R$, and $\Sigma^R$ in the GW cycle at $U/t=2$, as a function of the compression tolerance $\varepsilon$ from $10^{-3}$ to $10^{-12}$. The pole count grows approximately linearly in $\log_{10}(1/\varepsilon)$ for all three objects. At $\varepsilon=10^{-12}$, $A$ and $W^R$ require about 117 poles and $\Sigma^R$ about 55. We find that the self-energy is always the most compact object in this example, as it is smoother than $A$ and $W^R$ and has no sharp features at the band edges.

\begin{figure*}[t]
    \centering
    \includegraphics[width=1\linewidth]{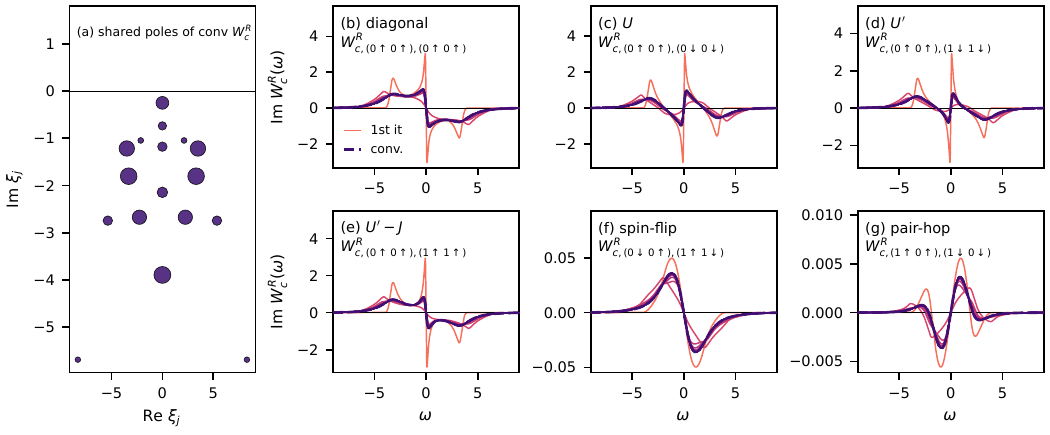} 
    \caption{Correlated matrix-valued screened interaction $\Im\,W_c^{R}(\omega)$
of the half-filled three-orbital Kanamori model ($U/t=1.6$, $J/U=0.2$) on a
Bethe lattice with half-bandwidth $D=2t$ at $\beta t=10$.
(a)~Shared pole locations $\xi_j$ of the converged $W_c^{R}$, with marker
area depicting the Frobenius norm of the residue matrices
$A_{W}\in\mathbb{C}^{N_{\mathrm{so}}^{2}\times N_{\mathrm{so}}^{2}}$ for the converged solution.
(b)--(g)~$\Im\,W_c^{R}(\omega)$ for representative matrix elements screening corresponding to diagonal  density-density,
 $U$, $U'$, $U'-J$, spin flip, and pair-hopping
channels. Light curves: iteration starting from the free Green's function. Dashed curve: converged solution.}
\label{fig:bethe_WR_Norb3}
\end{figure*}
Figure~\ref{fig:pole_budget}(b) shows how the pole counts evolve over the self-consistent iterations at tolerance $\varepsilon=10^{-8}$. We see that when the compression is applied based on a tolerance, $\mathcal{C}_\varepsilon$, the pole count remains adaptive unless a pole budget   $\mathcal{C}_M$ is deliberately imposed. If the dressing generates additional structure, this is reflected in the number of poles. This can be seen in the first few iterations where the self-consistency results in  the pole count to initially drift before settling to a nearly constant value for the remainder of the iterations. Figure~\ref{fig:pole_budget}(c) shows the converged $\Im\Sigma^R(\omega)/U^2$ for the same tolerances. It is an enlarged view of the GW self-energy of Fig.~\ref{fig:bethe_ImSigR_SelfConsistent}(c), and we focus on the region around its minimum, where the deviation between tolerances is large.  The curves for $\varepsilon\le10^{-6}$ are visually indistinguishable on the scale of the plot. Even at $\varepsilon=10^{-3}$, where $\Sigma^R$, $W^R$, and $A$ are represented by only 5, 10, and 15 poles, the self-energy remains qualitatively correct. The inset shows the deviation from the $\varepsilon=10^{-12}$ result and demonstrates that the error on the real axis is closely follows the compression tolerance set on the moment representation. The full self-consistent cycle therefore remains stable under a tight pole budget, and accuracy can be traded against computational cost in a controlled way. In our experience, a comparable error level in a Matsubara-axis calculation would likely degrade dynamical quantities far more severely.

\subsection{\label{ssec:MODMFT}Multi-orbital DMFT }
\emph{Model and setup.}
We now study a three-orbital model with degenerate orbitals
and a rotationally invariant Slater--Kanamori interaction~\cite{ImadaTokura1998}, using
GW as the impurity solver within DMFT. The model is a standard model for cubic
transition-metal oxides such as the titanates, vanadates and
ruthenates~\cite{Georges2013}. Within DMFT it has been studied extensively for spin freezing bad metal~\cite{Werner2008,Werner2009} and Hund's metal physics ~\cite{Werner2008,deMedici2011Janus,Mravlje2011Hund,Georges2013}, also
using numerically exact solvers~\cite{Werner2006,Gull2011}. While we do not expect the GW solver
to capture Mott physics, the model serves here as the simplest case
in which the screened interaction $W^R_c$ is a matrix-valued bosonic
object, and is therefore a test of the shared-pole representation.

The
local Hamiltonian is
\begin{align}
H_{\rm loc} &= -\mu\sum_{\alpha\sigma} n_{\alpha\sigma}
 + U\sum_{\alpha} n_{\alpha\uparrow}n_{\alpha\downarrow}
\nonumber\\
&\quad
 + \sum_{\alpha>\beta,\sigma}\Big[U' n_{\alpha\sigma}n_{\beta\bar\sigma}
 + (U'-J)\, n_{\alpha\sigma}n_{\beta\sigma}\Big]
\nonumber\\
&\quad
 - J\sum_{\alpha\neq\beta}\Big(
   c^{\dagger}_{\alpha\downarrow}c^{\dagger}_{\beta\uparrow}c_{\beta\downarrow}c_{\alpha\uparrow}
 + c^{\dagger}_{\beta\uparrow}c^{\dagger}_{\beta\downarrow}c_{\alpha\uparrow}c_{\alpha\downarrow}
 + {\rm h.c.}\Big).
\label{eq:kanamori}
\end{align}
Here $\alpha,\beta=1,2,3$ are the orbital indices, $\sigma=\uparrow,\downarrow$ is
the spin index, $n_{\alpha\sigma}=c^\dagger_{\alpha\sigma}c_{\alpha\sigma}$,
and $\mu$ is the chemical potential. $U$($U'$) is the intra-orbital
(inter-orbital) Coulomb interaction, and $J$ is the Hund's coupling interaction. The rotational invariance in the model sets $U' = U-2J$. We operate on the Bethe lattice with a semicircular density of states at half filling and $\beta t=10$. The bath,
$N_F$, $N_B$ and the mapping parameter $\omega_p$ are those of
Sec.~\ref{ssec:SSDMFT}, and the hybridization is updated as
$\Delta^R=t^2G^R$ after each impurity solve. All compressions use
$\varepsilon=10^{-8}$. We primarily study the weak coupling regime $U/t=1.6$ where the GW solution  reaches qualitative agreement with continuous time Monte Carlo \cite{Werner2006B}.

\emph{Shared poles.}
For $G$ and $\Sigma$, the shared-pole structure is trivial. Both objects
remain diagonal and identical for all spin-orbitals due to symmetry. The screened interaction $W^R_c$, however, is
matrix-valued in the bilinear basis $(ij)$ of spin-orbital pairs. Each
residue $A^W_j$ is therefore an $N_{\rm so}^2\times N_{\rm so}^2$
matrix while all matrix elements share the same set
of poles $\xi_j$. For diagonal $G$, the polarization satisfies
$\Pi_{(ij),(kl)}\propto\delta_{jk}\delta_{il}$, so the bubble couples
each channel $(ij)$ only to its conjugate $(ji)$. The interaction vertex
then generates channel mixing in $W^R_c$. The density
channels are coupled through $U$, $U'$ and $U'-J$. The spin-flip and
pair-hopping channels are coupled through $J$. The pole representation $\mathbf{W^R_c} = \mathcal{D}[U\mathbf{\Pi}^R]U$ of the screened interaction is obtained by diagonalizing the
upfolded matrix in Eq.~\ref{eq:upfolded_W} of the Dyson resolvent, which yields
$N_{\rm so}^2M_\Pi$ poles with rank-one residues given by
Eq.~\ref{eq:residue_W}. Consequently, it loses its compact structure. The compression $\mathcal{C}_\varepsilon \mathbf[{\mathbf{W^R_c}}] $
then gives a compact representation with few shared poles and full matrix-valued
residues.

\emph{Results.} Figure~\ref{fig:bethe_WR_Norb3}(a) shows the shared poles of the converged $\mathbf{W^R_c}$. The symbol area is chosen proportional to the Frobenius norm of the residue matrix $A^W_j$. Panels (b)--(g) show $\Im W^R_c$ for one representative of each inequivalent channel across the self consistent iterations. The first iteration is shown in the lightest shade and subsequent iterations in darker shades. 

Panel (a) demonstrates that all poles lie in the lower half-plane and satisfy the symmetry $\xi_j\mapsto-\xi_j$, $A^W_j\mapsto-[A^W_j]^\dagger$, which ensures the condition $W^R_c(-\omega)=W^R_c(\omega)^{\dagger}$. Exactly five poles lie on the imaginary axis and are fixed points of this symmetry; thus, their residues are purely imaginary. These poles generate an overdamped contribution $\propto\omega/(\omega^2+\gamma^2)$ to $\operatorname{Im}W^R_c$ at low frequency and dominate the static screening, $W^R_c(0)=-\sum_j A^W_j/\xi_j$. The residues also satisfy the sum rule $\left|\sum_j\operatorname{Tr}A^W_j\right|\lesssim10^{-7}$, within the specified compression tolerance. Both the sum rule and the pole symmetry can be enforced to higher precision by tightening this tolerance.

Panel (b) shows the screening in the diagonal equal-spin density--density channel. The Lehmann representation of the retarded density-density commutator requires the diagonal elements to satisfy the spectral-positivity condition $-\operatorname{sgn}(\omega)\operatorname{Im}W^R_c(\omega)\ge0$. Our results satisfy these constraints over the full frequency range at every iteration. The off-diagonal density channels in panels (c)--(e) similarly retain the required spectral antisymmetry, but are not required to be sign-definite. Their frequency dependence closely follows that of the diagonal channel. The spin-flip and pair-hopping components in panels (f) and (g) are two to three orders of magnitude smaller, consistent with their generation through Hund's coupling $J$ at second and third order, respectively. Finally, the sharp low-frequency features present in the first iteration are progressively smoothed by self-consistency, as in the single-orbital case of Sec.~\ref{ssec:MODMFT}.

\subsection{\label{ssec:UEG}The Uniform Electron Gas }
\emph{Model and setup.} As the final example, we consider the uniform electron gas (UEG) within the fully self-consistent GW approximation at finite temperature and revisit the satellite features. This model requires one to contend with momentum dependence. In essence, the continuous momentum requires one to work with a \emph{family} of scalar pole representations rather than one pole representation per object. Detailed results will be presented elsewhere (R. Farid {\it et al.}, unpublished). Here we demonstrate the main aspects.

The UEG enjoys rotational invariance, which reduces the momentum dependence to the radial coordinate $k=|\vec{k}|$. We work in Rydberg units, such that $\hbar=1$, $2m_e=1$, and $k_F=1$. The bare propagator is
$$
G_0^R(k,z)=\frac{1}{z-\epsilon(k)+i\eta},
\qquad
\epsilon(k)=k^2-\mu.
$$
We denote the Fermi wave vector by $k_F$, with density $n=k_F^3/(3\pi^2)$. The Coulomb interaction is $v(q)=4\pi g/q^2$, where $g=2r_s(4/9\pi)^{1/3}$ is set by the Wigner-Seitz radius $r_s$. Throughout, energies are measured in units of the Fermi energy, $\epsilon_F=1$, and radial momenta are cut off at $k_{\max}=4k_F$.

The free propagator is seeded with a small broadening $\eta$ to shift poles into the lower half-plane, as required for a retarded function. In a grid-based real-frequency scheme, an arbitrarily small $\eta$ would require a correspondingly fine frequency grid to resolve the resulting narrow Lorentzian in each frequency and momentum integral. In the present approach, $\eta$ simply shifts the pole position. In practice, we choose $\eta$ comparable to the compression tolerance, $\eta=\varepsilon=10^{-6}$ in this work.

The spin-summed $\Pi$ bubble and the $GW$ self-energy in the pole representation become
\begin{align}
\mathbf{\Pi}^{\lessgtr}_{q}&=-\frac{i}{2\pi^2 q}\int_0^{k_{\max}}\!\!dk\,k\int_{|k-q|}^{k+q}\!\!dp\,p\;\mathbf{G}^{\lessgtr}_{k}\ast\mathcal{I}\mathbf{G}^{\gtrless}_{p},
\label{eq:ueg_pi}\\
\mathbf{\Sigma}^{\lessgtr}_{c,k}&=\frac{i}{4\pi^2 k}\int_0^{k_{\max}}\!\!dq\,q\int_{|k-q|}^{k+q}\!\!dp\,p\;\mathbf{G}^{\lessgtr}_{p}\ast\mathbf{W}^{\lessgtr}_{c,q},
\label{eq:ueg_sc}
\end{align}
with $\mathbf{G}_k$ the pole representation at radial momentum $k$, and $\mathbf{W}^R_{c,q}=\mathcal{C}_\varepsilon\mathcal{D}[v(q)\mathbf{\Pi}^R_q]\,v(q)$ scalar in $q$. The chemical potential is readjusted at every iteration to fix $n$.

\begin{figure}
    \centering
\includegraphics[width=1\linewidth]{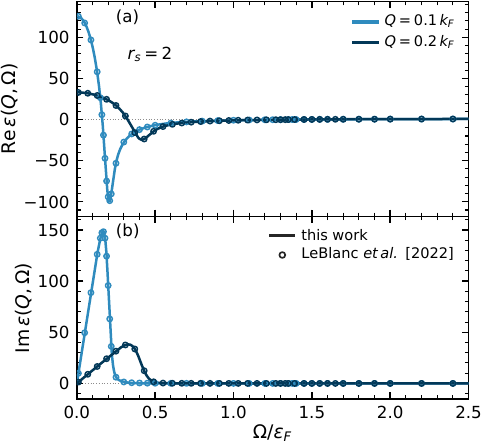}
    \caption{Dielectric function $\varepsilon(Q,\Omega)$ of the uniform electron
gas at $r_s=2$, $T/\varepsilon_F=0.1$, for $Q=0.1\,k_F$ and $0.2\,k_F$.
(a)~Real and (b)~imaginary parts for the free electron problem  benchmarked against Algorithmic Matsubara Integration (AMI) results . Solid lines: results from this work, evaluated directly
on the real frequency axis without analytic continuation. Circles:
results from Ref.~\cite{LeBlanc2022}, computed directly on the real axis with finite thermal broadening $\eta=\varepsilon_F/200$. In our work, we set $\omega \to \omega + 0.005i$  for direct
comparison. }
    \label{fig:UEF_benchmark}
\end{figure}

\emph{Moment interpolation} All many-body functions are computed on a discrete set of radial nodes $\{k_i\}\subset(0,k_{\max}]$, whereas the momenta $p$ in Eqs.~\eqref{eq:ueg_pi}--\eqref{eq:ueg_sc} range over a continuum rather than being restricted to these nodes. To perform these integrals one must therefore obtain a smooth representation of each object as a function of $k$ from its values at the nodes. The poles and weights of $\mathbf{G}_k$ do not interpolate well, as observed in \cite{Dong2026}.  Even when $G(k,\omega)$ changes smoothly with $k$, the poles can split and merge as $k$ varies,  and near a continuum they appear in clusters between which no correspondence exists.  The trigonometric moments $\{h_l^{G}(k)\}$ of Eq.~\eqref{eq:moments_poles}, on the other hand, are obtained from the function by the linear map Eq.~\eqref{eq:moments_quad} and therefore inherit its smooth $k$ dependence \cite{Dong2026}. The low moments are smooth in $k$ and vary slowly whereas higher moments vary more strongly and decay. We truncate the number of moments to 150 with conformal parameter $\oo_p=8$. These moments are stored on Chebyshev nodes in $k$. Whenever an intermediate value is required, it is obtained by Chebyshev interpolation of the moments followed by ESPRIT. The moment linearity is also exploited to perform momentum integrals directly in moment space  as discussed in Sec~\ref{sec:MinipoleOperations}.

\emph{Benchmark.} Fig.~\ref{fig:UEF_benchmark} panels (a,b) benchmark the real and imaginary components of the  dielectric function  $(1-\frac{4\pi g}{q^2}\Pi^R_{0})$   at $r_s=2$ and $T/\epsilon_{F}=0.1$ against Algorithmic Matsubara Integration (AMI \cite{Taheridehkordi2019AMI,Elzab2022, Burke}) results at $Q=0.1k_{F},0.2k_F$ obtained with  non-interacting propagators \cite{LeBlanc2022}. Here, we have set the compression parameter to  $\varepsilon=10^{-8}$. Note that Ref.~\cite{LeBlanc2022} computes $\Pi^{R}(Q,\omega) $ directly on the real axis with  a finite thermal broadening $\varepsilon_F/200$. Since our $\Pi^R_0$ is valid in the whole upper half plane, we evaluate it at $z=\omega+i\varepsilon_F/200$ for a direct comparison and find excellent agreement. 

\begin{figure}
    \centering
\includegraphics[width=0.95\linewidth]{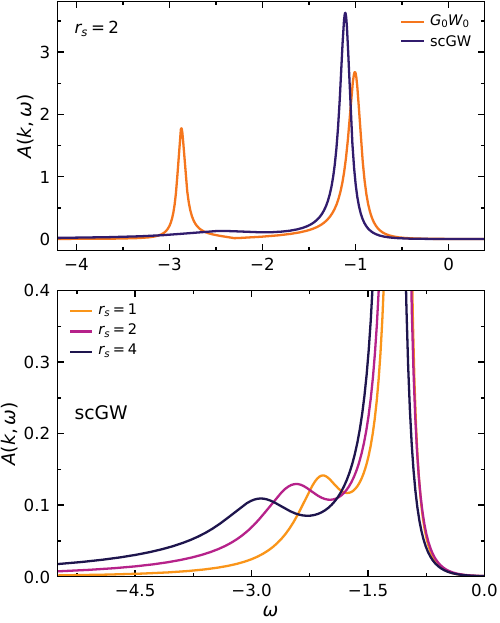}\\
    \caption{Top panel plots the spectral functions calculated within one-shot ($G_0W_0$) and fully self-consistent $GW$ (sc$GW$) for Wigner-Seitz radius $r_s=2$ near the bottom of the band at $T/\epsilon_{F}=0.10$. The bottom panel provides a zoomed-in view of the sc$GW$ spectral functions, highlighting satellite structure across all three densities.}
    \label{fig:UEF_rs_sat_zoom}
\end{figure}

\emph{GW results.} Figure~\ref{fig:UEF_rs_sat_zoom} shows $A(k,\omega)$ at the bottom of the band for $r_s=2$ from $G_0W_0$ and self-consistent $GW$ , and the satellite region of the $GW$ spectral function for $r_s=1,2,4$. The calculations were carried out  at  $T/\epsilon_{F}=0.1$ and at a fixed density $n=1/3\pi^2$. The $G_0W_0$ calculations exhibit a sharp peak (a `plasmon pole') below the dominant quasi-particle peak. In the pole representation, both  features appear directly as the dominant poles and weights. The $G^R$ propagator in the $G_0W_0$ calculation is dominated by two poles that correspond to nearly $98\%$ of the weight. The quasi-particle peak is carried by the pole $\xi_1^G=-1.00-0.08i$, and the satellite is carried by the pole $\xi_2^G=-2.88-0.05i$.  The real part of each pole gives the frequency of the feature and the imaginary part its width, i.e.~the inverse lifetime of the excitation. The small imaginary part of $\xi_2^G$ is what makes the $G_0W_0$ satellite sharp. The contribution to spectral weight is obtained from  $\Re A^G_j$ with spectral sum  obeying ($\sum_{j} \Re A^G_j$=1). In $G_0W_0$, we obtain  $\Re A^G_1=0.70$ for the quasi-particle pole and $\Re A^G_2=0.28$ in the satellite pole.  Under self-consistency the quasi-particle pole moves to $\xi^G_1=-1.11-0.07i$ and its weight increases to $ \Re A^G_1=0.80$, while the satellite pole moves to $\xi^G_2=-2.47-0.64i$ with weight $\Re A^G_{2} = 0.19$. Therefore, we find the satellite position is shifted, its damping increased by an order of magnitude, and its weight reduced. Nevertheless, the satellite features do not disappear. Rather, although broadened by self-consistency, the satellite remains a well-defined feature at all three densities. 

This differs from the fully self-consistent calculation of Holm and von Barth~\cite{HolmVonBarth1998}, who found the satellite to be washed out into a broad, featureless background. In that work the spectral function was represented as a sum of Gaussians, and a fit of this form may have smoothed a weak but well-defined feature. Our results, in which the frequency dependence is treated exactly up to the compression tolerance, indicate that a satellite with a broadened maximum survives full self-consistency, but with a marked reduction in weight and increased damping relative to $G_0W_0$. This is consistent with the recent finding of Ref.~\cite{Harsha2024}, who observed with accurate imaginary-time grids and tight convergence that $GW$ retains the plasmon satellites of diamond with reduced weight, where earlier calculations had lost them.

\section{Discussion}\label{sec:discussion}
In the following, we introduce several aspects of pole diagrammatics with the aim of helping a reader connect to other diagrammatic theories or other aspects of Green's function methodology. We first connect to the DLR  \cite{DLR_original_PhysRevB.105.235115} and the sum-over-poles representation \cite{ChiarottiMarzariFerretti2022} in more detail, before introducing the hyperfunction picture of Green's functions \cite{smit2022hyperfunctionformulationbodygreens}, and  outlining connections to quasiparticles and the plasmon pole approximation.
\begin{figure}[tb]
    \centering
    \includegraphics[width=1.0\linewidth]{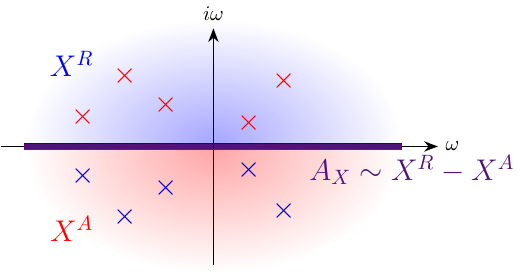}
    \caption{A pair of analytic functions sharing a boundary defines a hyperfunction, encoded by their difference at the boundary. The two functions $X^{R/A}(z)=\mathcal{S}A_X(z\pm i0^+)$ are analytic on $\mathbb{C}^{+}$ and $\mathbb{C}^{-}$, respectively, and their boundary difference defines the spectral function $A_X(\omega)$ on the real line.}
    \label{fig:minimalpole_data_entire_complex_plane}
\end{figure}

\emph{The Discrete Lehmann representation.} The most widely used pole representation in correlated and dynamical mean field calculations is the Discrete Lehmann Representation (DLR)~\cite{DLR_original_PhysRevB.105.235115,Kaye2024_DLR_cpplib}, which emerged after a series of earlier attempts to find compact representations of imaginary time and Matsubara Green's functions \cite{BoehnkeHafermann2011,Kananenka2016,ShinaokaOtsuki2017,Gull2018,Li2020}. In DLR, Matsubara functions are represented as a sum over poles on the real axis. The pole locations are function and problem independent: for a given energy window $\Lambda$, inverse temperature $\beta$ and target tolerance $\varepsilon$, the DLR generates a fixed set of poles that can approximate any Matsubara Green's function at the Matsubara frequency points within $\varepsilon$. This universality is particularly convenient for performing Matsubara and momentum sums such as those arising in Dynamical Mean-Field Theory (DMFT) and higher-order diagrammatics, as these reduce to a sum over the pole weights for known pole locations \cite{Gazizova2024,Gazizova2025}. However, the DLR representation is only accurate on the Matsubara points, necessitating numerical analytic continuation methods if values on the real axis or between Matsubara points are needed.

In contrast to the fixed set of DLR frequencies valid for all points, our minimal pole approach provides a minimal and optimal set of complex poles tailored to a specific Green's function that is valid on the entirety of the complex plane. For momentum sums, our approach leverages its associated moment representation over which summations extend linearly.

\emph{The Sum-Over-Poles representation.} Another approach that resembles aspects of the minimal pole approach is the Sum-Over-Poles (SOP) approach~\cite{ChiarottiMarzariFerretti2022,ChiarottiFerrettiMarzari2024}. SOP also employs non-universal complex-valued poles and weights to efficiently evaluate Dyson resolvents and perform algebraic operations without frequency grids. Our  approach presents two key conceptual advances. First, in the SOP approach, the pole sums are re-expressed as a sum over generalized $n$-th order Lorentzians, and their parameters are subsequently fit with a non-linear fitting scheme. In our approach, the fitting problem for the poles and weights is recast into a Prony-type approximation problem, for which robust signal processing schemes like ESPRIT are used to determine the optimal set of poles. This utilizes an intermediate moment representation, and is made stable with a conformal mapping of the complex poles $\xi_j$ into the unit disk such that $|\xi_j| < 1$. Second, our approach introduces a convenient moment representation that enables seamless re-compression operations as well as momentum summations and interpolation, bypassing the need for frequency grids.

\emph{Errors away from the real axis.} Given a Green's function $X(z)$, we typically construct a compact pole representation for its retarded piece $\mathbf{X^R}$. Because $\mathbf{X^R}$ consists of poles solely in the LHP, the error $\Delta X^R(z) = \mathbf{X^R}(z) - X^R(z)$ is complex analytic in the UHP and vanishes at infinity.  If the approximation agrees with the retarded function $X^R(z)$ on the real axis up to a tolerance $\varepsilon$, the maximum modulus principle  guarantees that the bound $|\Delta X(z)| \leq \varepsilon$ holds for $\Im z > 0$. Similar arguments follow for the advanced function.

\emph{The hyperfunction picture on the entirety of the complex plane.} The view of two `sheets' in the complex plane (in analogy to the logarithm or square root)  is useful in this context. The pair $(\mathbf{X^R}, \mathbf{X^A})$ provides a compact representation for $\mathcal{S}A_X(z)$ on the split complex plane $\mathbb{C}\setminus \mathbb{R}$, if one identifies its value in the UHP as given by $\mathbf{X^R}$ and in the LHP as given by $\mathbf{X^A}$. Their boundary difference along the real axis accurately captures the discontinuity given by the spectral function $A_X$. $\mathbf{X^R}$ therefore provides a compact representation of $\mathcal{S}A_X$ in the sense of hyperfunctions~\cite{smit2022hyperfunctionformulationbodygreens}, as is illustrated in  Fig.~\ref{fig:minimalpole_data_entire_complex_plane}.

\emph{Quasi-particle residues.} Because the minimal pole approach provides a representation over a continuous complex domain, the quasi-particle residue $Z=(1-\partial \Re \Sigma^R(\omega)/\partial \omega|_{\omega=0})^{-1}$ can be calculated directly by analytically evaluating the derivative to obtain $Z=\left[1+\sum_{j=1}^M {A^\Sigma_j}/{(\xi_j^{\Sigma})^2}\right]^{-1}$. In contrast, standard Matsubara approaches evaluate this quantity using the finite difference estimate $\zeta(T)=[1-\text{Im} \Sigma(i\omega_0)/(i\omega_0)]^{-1}$, where $\omega_0=\pi T$ and $T$ is the temperature~\cite{GeorgesKotliar1996}. Although this expression is formally exact as $T \rightarrow 0$, it breaks down at finite $T$ when the self-energy develops sharp structures at frequencies smaller than $\pi T$.

The {\it Plasmon Pole} approximation (PPA) \cite{Lundqvist19671} is frequently used in GW calculations of real materials when the full frequency-dependence of the correlated screened interaction is not required.
The PPA can be understood as a pole expansion in the sense of this work, where the expansion is truncated to a single pole at positive frequency and its conjugate at negative frequency~\cite{Bonacci2026,Hwang2018}.
In practice, there is considerable freedom in how precisely the pole frequency and residue are chosen ~\cite{Larson2013,Shih2010,Engel1993,Linden1988}. A popular and historically significant approach consists of fixing them by the static dielectric matrix together with a generalized
$f$-sum rule~\cite{Hybertsen1986}. An alternative reproduces the dielectric matrix at
two a priori chosen frequencies, $\omega=0$ and an imaginary frequency
$\mathrm{i}\omega_{\mathrm{fit}}$~\cite{Godby1989}.
Since both $G_{0}$
and the model $W^{c}$ are then rational functions of frequency, the $GW^{c}$
convolution reduces to residues and only one or two dielectric matrices must be
constructed and stored, which is why the PPA remains a standard production workflow
for large systems.
The extent to which the truncation to a single pole reduces the accuracy of real-world calculations is a topic of research~\cite{Aryasetiawan1998,Leon2021}, and a truncation of our scheme to $M=1$ for $W_c$ has not yet been attempted.

\section{Conclusions}\label{sec:conclusions}
In conclusion, we have introduced a formulation of real-frequency diagrammatics in terms of moment and pole representations that is well suited to high-precision calculations of spectral quantities. We have documented the diagrammatic operations needed for single-shot and self-consistent diagrammatics, and we have introduced a scheme that avoids the proliferation of poles that plagued earlier implementations of such schemes. Our representations are controlled and systematically improvable in practice, and sufficiently compact to allow for efficient calculations of complex quantum systems. Shared pole representations allow the efficient simulation of multi-orbital quantities.

The methodology is entirely free of grid discretization approximations and their associated errors, and numerically efficient due to the low-information structure of the pole representations.

Pole approximations of this type open an avenue towards high-precision simulation of spectral quantities of quantum systems at non-zero temperature. This work presented results for simple model systems along with a simulation of the uniform electron gas within self-consistent $GW$. Low-order perturbative simulations of realistic systems such as those currently performed with Matsubara codes \cite{Yeh2022,Iskakov2024} as well as higher-order diagrammatics \cite{Taheridehkordi2019AMI} with efficient decomposition schemes \cite{KayeHuangStrandGolez2024,NunezFernandez22,Erpenbeck23,Yeh2024} or vertex corrections \cite{Stefanucci2014,Pavlyukh2016,Pokhilko2024,Pokhilko2025} are obvious next steps. Extending the framework to diagrammatic formulations beyond the equilibrium Keldysh framework and to applications outside condensed-matter physics also offers promising directions for future research.

\begin{acknowledgments}
We thank JPF LeBlanc for helpful and productive discussions.
This work has been sponsored by the European Research Council (ERC) under Advanced Grant No. 101142136 (Quantum
Algorithms).
\end{acknowledgments}

\section*{Data Availability}
Simulation data and computer codes are available from the
authors upon reasonable request.

\appendix

\section{The ESPRIT algorithm}\label{app:esprit}
In this appendix we briefly describe the numerical algorithm used to obtain the pole expansion of Eq.~\ref{eq:moments_poles}.
In general, given \(N\) values of a function \(f(t)\) sampled on a uniform grid \(t_k\),
\(0 \leq k \leq N-1\), and a tolerance \(\epsilon\), a {\it Prony approximant} yields the minimal number \(M\) of complex nodes \(z_j\) and complex residues \(R_j\) such that
\begin{align}\label{eq:prony_problem_reformulate}
    \big|f(t_k) - \sum_{j=1}^M R_j z_j^{k}\big| \leq \epsilon,
    \quad \text{for all } 0 \leq k \leq N-1 \; .
\end{align}
In the present application, the samples \(f(t_k)\) correspond to the moments \(h_k\) of Eq.~\ref{eq:moments_poles}, the residues \(R_j\) to the transformed weights \(A_j'\), and the nodes \(z_j\) to the corresponding poles.

We solve Eq.~\ref{eq:prony_problem_reformulate} with the ESPRIT algorithm, which we now outline. The method is based on a singular value decomposition (SVD) of the \((N-L) \times (L+1)\) Hankel matrix
\begin{equation}\label{eq:hankel_matrix}
    H =
    \begin{pmatrix}
        f(t_0)       & f(t_1)     & \cdots & f(t_{L}) \\
        f(t_1)       & f(t_2)     & \cdots & f(t_{L+1}) \\
        \vdots    & \vdots  & \ddots & \vdots \\
        f(t_{N-L-1}) & f(t_{N-L}) & \cdots & f(t_{N-1})
    \end{pmatrix}
\end{equation} 
expressed as  
\begin{equation}
    H = U \Sigma W \;,
\end{equation}  
where \( U \) and \( W \) are unitary matrices of dimensions \((N-L) \times (N-L)\) and \((L+1) \times (L+1)\), respectively. The matrix \( \Sigma \) is an \((N-L) \times (L+1)\) rectangular diagonal matrix, with its diagonal entries arranged in descending order, \(\sigma_1 \geq \sigma_2 \geq \cdots \geq \sigma_{L+1} \geq 0\). In practice, \(L\) is typically chosen within the range \(N/3 \leq L \leq N/2\) to minimize the variance~\cite{Sarkar1995}, and we set \( L = 2N/5 \)  throughout our implementation unless otherwise stated.


For a target tolerance \( \epsilon \), the number of exponentials \( M \) is estimated as the smallest index satisfying the absolute tolerance criterion of \( \sigma_{M+1} \leq \epsilon \) . This truncation yields the minimal number of exponentials required to satisfy the prescribed error tolerance. The nodes \(z_j\) are then obtained as the eigenvalues of the matrix
\begin{equation}
    F = (W_1^T)^+ W_2^T \; ,
\end{equation}
where \(T\) denotes the transpose and \(+\) the pseudoinverse. The matrices \(W_1\) and \(W_2\) are constructed from \(W\) according to
\begin{equation}\label{eq:esprit_Wm}
    W_s = W(1:M, s:L+s-1), \quad s = 1,2,
\end{equation}
where $W_1$ and $W_2$ are obtained from the first $M$ rows of $W$ by deleting the last and the first column, respectively. Finally, the weights \(R_j\) are computed by solving the following overdetermined Vandermonde system in the least-squares sense:
\begin{equation}\label{eq:solve_weights_matrix}
\begin{pmatrix}
    f(t_0) \\
    f(t_1) \\
    \vdots \\
    f(t_{N-1})
\end{pmatrix}
\!=\!
\begin{pmatrix}
    1 & 1 & \cdots & 1 \\
    z_1 & z_2 & \cdots & z_M \\
    \vdots & \vdots & \ddots & \vdots \\
    z_1^{N-1} & z_2^{N-1} & \cdots & z_M^{N-1}
\end{pmatrix}\!\!
\begin{pmatrix}
    R_1 \\
    R_2 \\
    \vdots \\
    R_M
\end{pmatrix}.
\end{equation}
The resulting set of nodes and weights provides a compact exponential representation that reproduces the sampled data. Due to the effect of the non-linear Mobius transform, the number of poles and residues kept may need to be adjusted to ensure an approximation of the original data to within \(\epsilon\).

For multi-orbital systems, computational efficiency can be improved by employing a generalization of ESPRIT for matrix-valued functions~\cite{ZhangYuGull2024}, in which the matrix elements share a common set of nodes while retaining matrix-valued weights. This approach follows a similar idea to that introduced in Ref.~\cite{YingPoleRecovery2022} as a matrix-valued extension of the scalar Prony method developed in Ref.~\cite{YingAC2022}.

\section{Benchmarks of the Padé spectral decomposition in convolution integrals}
\label{app:pade_spectral}
\begin{figure}[tb]
\centering
\includegraphics[width=1\linewidth]{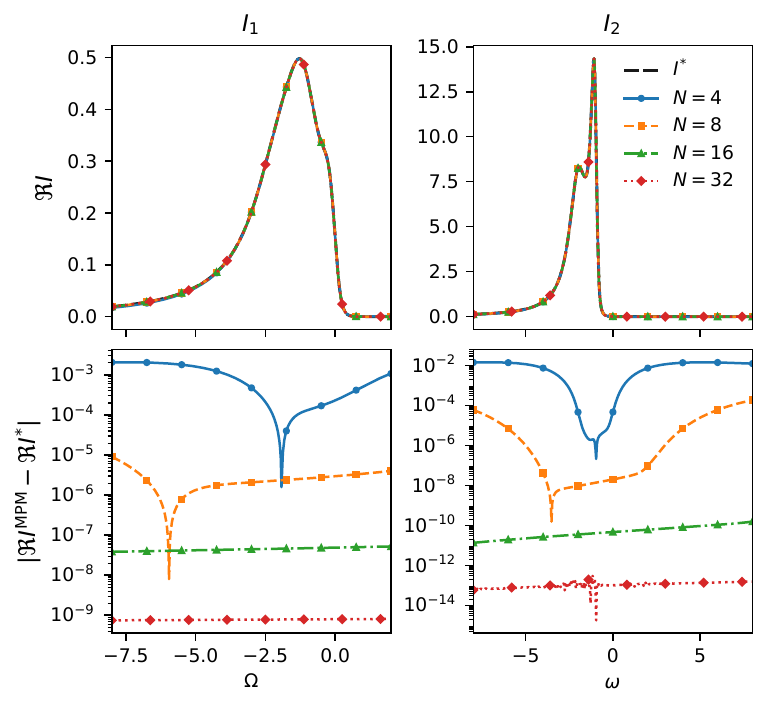}
\caption{Convergence of the convolution integrals $I_1$ (first column) and $I_2$ (second column) at $\beta=10$ as the number of positive-imaginary PSD poles $N$ for the Fermi and Bose distribution functions is increased. Results are compared to numerical quadrature.}
\label{fig:I12_benchmark}
\end{figure}

Here we give the details of the two convolution integrals $I_1$ and $I_2$ benchmarked in Fig.~\ref{fig:I12_benchmark}. For the fermionic bubble $I_1$, we consider a synthetic spectral function that admits a three peak structure, which we model as a sum of Lorentzians
\begin{align}
A_0(\omega) = \frac{1}{3} 
\sum_{\epsilon =-1}^1 \frac{1}{\pi} \frac{\gamma_{\epsilon}}{(\omega-\epsilon)^2 + \gamma^2_\epsilon},
\end{align}
 with $\gamma_{-1} = \gamma_1 = 0.6, \gamma_0 = 0.2.$ The bubble integral reads
\begin{align}
I_1(\Omega) = -\int_{-\infty}^\infty \frac{\dd{\omega }}{2\pi}  g_0^<(\omega) g_0^>(\omega - &\Omega),
\end{align}
where $g_0$ represents the Green's function corresponding to $A_0$.
From Eq.~\ref{eq:FDT}, we obtain
\begin{align}
 I_1 = \int_{\mathclap{-\infty}}^{~\mathclap{\infty}} 
 \dd{\omega} 2\pi n_F(\omega) A_0(\omega)  (1-n_F(\omega-\Omega))A_0(\omega-\Omega).
\end{align} 
We use inverse temperature $\beta = 10$, at which the expression can be evaluated to very high accuracy with standard numerical quadrature on a grid.  To carefully benchmark the performance of the MPM combined with the PSD, we rewrite this expression in the pole representation using the algebraic operations defined in Sec.~\ref{sec:MinipoleOperations}. The spectral function $A_0$ we have chosen admits an explicit pole representation with $6$ poles. The part of the Fermi-distribution function $\tilde{n}_F = n_F - 1/2$ is approximated by $M=2N$ PSD poles ($N$ positive imaginary poles). The MPM expression is then given by
\begin{align}
\mathbf{I_1} = (2\pi)^2\Big(\frac{1}{4}\mathbf{A_0}\ast \mathcal{I}\mathbf{A_0} +\frac{1}{2}(\mathbf{\tilde{n}_F}\mathbf{A_0})\ast \mathcal{I}\mathbf{A_0} - \nonumber \\ \frac{1}{2}\mathbf{A_0}\ast \mathcal{I}(\mathbf{\tilde{n}_F}\mathbf{A_0}) - (\mathbf{\tilde{n}_F}\mathbf{A_0})\ast \mathcal{I}(\mathbf{\tilde{n}_F}\mathbf{A_0})\Big). \label{eq:I1_integral}
\end{align}
 In the first column of Fig.~\ref{fig:I12_benchmark}, we compare a quadrature evaluation of Eq.~\ref{eq:I1_integral} denoted by $I^*_1$ with an absolute error of up to $10^{-13}-10^{-14}$, extracted by varying the number of grid points. We benchmark the pole calculation for various $N$. We find the error in the integral to systematically decrease with $N$, and already at $N = 32$ positive imaginary poles, we can reach errors reaching roughly $10^{-9}$ over most of the frequency window. 

For the second integral $I_2$, the damped bosonic mode has spectral function 
$$B_0(\Omega) = 2\eta\left[\frac{1}{(\Omega-\epsilon)^2+\eta^2} - \frac{1}{(\Omega + \epsilon)^2 + \eta^2}\right],$$ with energy $\epsilon= 1$ and damping parameter $\eta = 10^{-2}$. 
The integral reads
\begin{align}
I_2(\omega) &= \int_{-\infty}^\infty \frac{\dd{\Omega}}{2\pi} d^<_0(\Omega)g_0^<(\omega-\Omega),
\end{align}
where $d_0$ is the bosonic Green's function corresponding to $B_0$. Defining $\tilde{n}_B = n_B + \frac{1}{2}$, and utilizing Eqs.~\ref{eq:FDT}, the expression can be written in the pole representation as
\begin{align}
\mathbf{I_2} = (2\pi)^2  \big(   -\frac{1}{4}\mathbf{B_0}\ast \mathbf{A_0}+\frac{1}{2}(\mathbf{\tilde{n}_B}\mathbf{B_0})\ast \mathbf{A_0}   -\nonumber \\ \frac{1}{2}\mathbf{B_0}\ast(\mathbf{\tilde{n}_F} \mathbf{A_0})+ (\mathbf{\tilde{n}_B}\mathbf{B_0})\ast(\mathbf{\tilde{n}_F} \mathbf{A_0})\big).
\end{align}
There is a subtlety in calculating this convolution of pole representations, namely the presence of a pole of $\mathbf{\tilde{n}_B}$ at zero which gives an ambiguous evaluation for the  $
\Theta$-function in Eq.~\ref{eq:convolution}. This ambiguity corresponds to the need of prescribing whether the pole at zero should lie inside or outside a semi-circular integration contour, or whether it should be treated in a principal-value sense. For our purposes, we need not be concerned with this, as the Bose function appears (and generally often does) next to a bosonic spectral function which vanishes at $z = 0$. Consequently, the result is independent of how we evaluate the $\Theta$-function, and considerations of this pole at zero can  safely be ignored. The Fermi and Bose functions are approximated within their PSDs to $M = 2N$, $M = 2N + 1$ poles, respectively. The bosonic spectral function $B_0$ admits an exact pole representation with 4 poles. We benchmark again  against quadrature evaluation of the integral on a frequency grid with up to $10^{-12}$ error, which we denote by $I^*_2$. We find systematic and rapid convergence with increasing $N$ as shown in the second column of Fig.~\ref{fig:I12_benchmark}.

 \section{Pole representation of the Dressing Operator} \label{app:dressing_operator}
 In this appendix, we describe how the dressing operator is implemented directly on pole representations. 
 Consider a $n \times n$-matrix valued kernel $K(z) + K_\infty$, where $K_\infty$ is a constant, while $K(z)$ decays to zero at $|z| \rightarrow \infty$ and admits a pole representation. In the solution of Dyson-type equations, it is desired to calculate
 \begin{align}
 \mathcal{D}(K(z)+K_\infty) = (\mathbbm{1} - K(z)-K_\infty)^{-1} - (\mathbbm{1}-K_\infty)^{-1},
 \end{align}
 with $\mathbbm{1}$ the $n\times n$ unit matrix, 
 directly in the pole representation. Note that the result decays to zero as $|z| \rightarrow \infty$. We denote this map by
 \begin{align}
 \mathcal{D}_{K_\infty}: \mathbf{K} \mapsto \left((\xi^\mathcal{D}_k)_k, (A_k^{\mathcal{D}})_k\right),
 \end{align}
 or simply $\mathcal{D}$ without the subscript when $K_\infty = 0$ as occurs in the physical case of the Dyson equation for the electron propagator or the screened interaction. The poles and weights are determined as follows. The poles $\xi^\mathcal{D}_j$ are the roots
 \begin{align}
 \operatorname{det}(\mathbbm{1} - K(\xi)-K_\infty) = 0
 \end{align}
 which can be obtained using a generalization of the companion matrix method~\cite{alanedelman1995}: the roots are the eigenvalues of the so-called companion matrix given by
 \begin{align}
 H
 =
 \begin{pmatrix}
 \xi^{K}_1\mathbbm{1} + T A^{K}_1 & \cdots &  T A^{K}_{M_K} \\
 \vdots & \ddots & \vdots \\
  T A^{K}_1 & \cdots & \xi^{K}_{M_K}\mathbbm{1} +T A^{K}_{M_K}
 \end{pmatrix},
 \label{eq:pole_representation_dressing_companion}
 \end{align}
 where $T := (\mathbbm{1} - K_\infty)^{-1}$, 
 and $M_K$ is the number of poles in the pole representation $\mathbf{K}$ of $K$. The corresponding weights follow from taking the residues at each simple root,
 \begin{align}
 A^{\mathcal{D}}_{k}
 =
 \frac{v_k w_k^\dagger}{-\,w_k^\dagger K'(\xi^{\mathcal{D}}_{k})v_k}
 = \frac{v_k w_k^\dagger} {w_k^\dagger
 \left(
 \sum_{m}
 \tfrac{A^{K}_{m}}{(\xi^{\mathcal{D}}_{k}-\xi^{K}_{m})^{2}}
 \right)^{-1}v_k},
 \label{eq:pole_representation_dressing_D}
 \end{align}
 with block ($n$-dimensional) null vectors
 \begin{align}
 v_k = \sum_{m=1}^{M_K} R_{k,m}, \quad w_k = T^\dagger \sum_{m=1}^{M_K} L_{k, m},
 \end{align}
 where $L_{k}^\dagger$ and $R_{k}$ are the left and right eigenvectors of $H$ respectively corresponding to $\xi^\mathcal{D}_k$, with $R_{k,m}$ referring to the vector acting on a sub-block $m$ of $H$. Assuming the poles remain simple after applying the dressing operator, $A_k^\mathcal{D}$ will be rank-1. In general, applying the dressing operator to pole representation with rank $M_K$ can yield $n M_K$ poles.
 Importantly, the dressing operator replaces an otherwise infinite geometric series of products by an eigenvalue problem of an $n M_K \times n M_K$ matrix.

\bibliographystyle{apsrev4-2}
\bibliography{refs}

\end{document}